\documentclass[aps,prb,notitlepage,reprint,floatfix]{revtex4-1}
\usepackage{graphicx,dcolumn,bm,amsmath,amssymb,euscript}
\usepackage[utf8]{inputenc}
\usepackage[T1]{fontenc}
\usepackage[colorlinks,citecolor=red,urlcolor=blue]{hyperref}
\usepackage[update,prepend]{epstopdf}       
\usepackage{color, colortbl}
\pdfoutput=1
\usepackage{pdfpages}  
\definecolor{LightCyan}{rgb}{0.88,1,1}
\graphicspath{{figures/}}

\begin{document}

\title{Kinetic Monte Carlo simulations of GaN homoepitaxy on c- and m-plane surfaces}

\author{Dongwei Xu}
\email[correspondence to: ]{xud@anl.gov}
\author{Peter Zapol}
\author{G. Brian Stephenson}
\affiliation{Materials Science Division, Argonne National Laboratory, Argonne, IL 60439}
\author{Carol Thompson}
\affiliation{Department of Physics, Northern Illinois University, DeKalb IL 60115}

\date{November 8, 2016} 

\begin{abstract}

The surface orientation can have profound effects on the atomic-scale processes of crystal growth, and is essential to such technologies as GaN-based light-emitting diodes and high-power electronics. We investigate the dependence of homoepitaxial growth mechanisms on the surface orientation of a hexagonal crystal using kinetic Monte Carlo simulations. To model GaN metal-organic vapor phase epitaxy, in which N species are supplied in excess, only Ga atoms on a hexagonal close-packed (HCP) lattice are considered. The results are thus potentially applicable to any HCP material. Growth behaviors on c-plane ${(0 0 0 1)}$ and m-plane ${(0 1 \overline{1} 0)}$ surfaces are compared. We present a reciprocal space analysis of the surface morphology, which allows extraction of growth mode boundaries and direct comparison with surface X-ray diffraction experiments. For each orientation we map the boundaries between 3-dimensional, layer-by-layer, and step flow growth modes as a function of temperature and growth rate. Two models for surface diffusion are used, which produce different effective Ehrlich-Schwoebel step-edge barriers, and different adatom diffusion anisotropies on m-plane surfaces. Simulation results in agreement with observed GaN island morphologies and growth mode boundaries are obtained. These indicate that anisotropy of step edge energy, rather than adatom diffusion, is responsible for the elongated islands observed on m-plane surfaces. Island nucleation spacing obeys a power-law dependence on growth rate, with exponents of -0.24 and -0.29 for m- and c-plane, respectively. 

\end{abstract}


\maketitle

\section{Introduction}

GaN-based semiconductors are widely used for optoelectronic devices\cite{Nanishi2014} and are being developed for high power applications.\cite{Nie2014} Typically wurtzite GaN films for these applications are grown in the ${(0 0 0 1)}$ c-plane orientation. While it was discovered early how to grow high quality films in the c-plane orientation, there has been much progress recently in growth on other surface orientations that offer potential advantages. For light emitting devices, the intrinsic electric field associated with polar c-plane orientations can limit performance,\cite{DenBaars2013,Waltereit2000} and use of non-polar orientations such as ${(0 1 \overline{1} 0)}$ m-plane can alleviate this problem.\cite{Nakamura2013} For high power devices, use of a vertical device geometry involving growth on various surface planes can improve performance.\cite{Nie2014,Fujit2015} This motivates our effort to model GaN growth on different crystal surface orientations such as c- and m-plane to elucidate the influence of orientation on growth modes and kinetics.  

In this study, we use kinetic Monte Carlo (KMC) simulations to observe the effects of surface orientation on atomic-scale mechanisms occurring during homoepitaxy of GaN films by typical methods, such as metal-organic vapor phase epitaxy (MOVPE). Several previous KMC studies of GaN growth have been carried out for the ${(0 0 0 1)}$ c-plane surface. Studies modeling GaN growth by molecular beam epitaxy \cite{Chugh2015,Wang1994} proposed mechanisms by which Ga and N diffusion rates and the Ga/N supply ratio influence surface morphology. A previous KMC simulation describing MOVPE \cite{Fu2008} focused on the chemical reaction, adsorption, and desorption processes without considering surface kinetics such as diffusion of atoms, attachment at step edges, etc. A sequence of studies that focus on surface kinetics \cite{ Zaluska_Kotur2012, Zaluska_Kotur2011, Zaluska_Kotur2010} have modeled step morphologies and instabilities as a function of growth conditions. The effect of the Ehrlich-Schwoebel step-edge barrier \cite{ehrlich1966atomic,schwoebel1966step} on step instabilities \cite{ Zaluska_Kotur2012} and growth mode transitions \cite{Kaufmann2016} has been studied. Recent KMC studies \cite{Krzyzewski2016,Krzyzewski2016a} have investigated step instabilities on the ${(0 0 0 \overline{1})}$ surface. Here we develop a model to compare the growth of GaN on two different surface orientations, ${(0 0 0 1)}$ c-plane and ${(0 1 \overline{1} 0)}$ m-plane, to observe effects of the crystal lattice structure on atomic-scale mechanisms determining homoepitaxial growth mode boundaries as a function of temperature and growth rate.

We analyze the surface structures observed in reciprocal space, to determine growth mode boundaries and mean island spacings, and to make contact with \emph{in situ} surface X-ray and electron scattering studies. Such experiments provide quantitative characterization of atomic-scale surface morphology during growth. In particular, X-ray methods can penetrate the MOVPE environment to reveal growth behavior as a function of conditions. We compare our KMC simulation results with X-ray studies of GaN MOVPE \cite{StephensonAPL1999,ThompsonJECS, MunkholmAPL, Perret2014} to fix the relationship between simulation and experimental timescales and provide physical insight into observed behavior.

The plan of this paper is as follows. In section II, we describe the features of our KMC model, and their direct implications for step edge energies and diffusion barriers on GaN c- and m-plane surfaces. In section III, we present growth simulations as a function of temperature and growth rate, and analyze the structures in reciprocal space to obtain island spacings and growth mode boundaries. In section IV, we compare the results with experiments on GaN MOVPE, and in section V discuss results and conclusions. Note that the results of the simulations can also be applied to other hexagonal materials by alternative choices of scaling parameters.

\section{KMC Model for {GaN} MOVPE}

Diffusion and chemical reactions of precursor species on the GaN surface under MOVPE conditions are not well understood. Indirect estimates of surface transport rates and mechanisms have been made, e.g. based on the observed temperature and length scale dependence of surface smoothing. \cite{2014_Koleske_JCrystGrowth391_85} A number of density functional theory calculations for diffusion and reaction energies and barriers of Ga and N species including ammonia have been reported, providing insights into molecular mechanisms of the growth processes.\cite{Lymperakis2009, Krukowski2009, 2012_Walkosz_JChemPhys137_054708, 2012_Walkosz_PRB_85_033308, An15, Jindal2010} In general, KMC calculations can take advantage of parameter values based on such DFT results. However, in the absence of a complete set of reliable parameters for different GaN orientations, we have been using a more generic energy model that we relate to experimental studies. 

While a full atomic-scale description of all processes occuring during MOVPE is challenging, one aspect of GaN growth provides a simplification. Because of the very high equilibrium vapor pressure of N$_2$ at the Ga/GaN phase boundary at typical growth temperatures,\cite{Ambacher1996} the nitrogen precursor (e.g. NH$_3$) is typically provided in large excess. Thus the surface is saturated with respect to nitrogen species (e.g. NH$_x$) in a dynamic steady state environment, \cite{2012_Walkosz_JChemPhys137_054708,2012_Walkosz_PRB_85_033308} and the rate-limiting steps for growth involve the deposition and incorporation of Ga.\cite{Perret2014} In our model we can thus focus on the behavior of Ga atoms, and assume the N structure remains in local equilibrium with the environment. A version of this assumption has been used in previous simulations. \cite{Zaluska_Kotur2012, Zaluska_Kotur2011, Zaluska_Kotur2010} While considering only the Ga sites involves an approximation, it allows us to investigate the anisotropies on various crystal faces due to the primary underlying crystal symmetry. 

\subsection{Choice of HCP lattice for simulation}

To simulate MOVPE of GaN, we use a KMC model based on a crystal lattice of Ga atomic sites, where each site is either occupied by an atom or not. The Ga sites in the wurtzite GaN structure form a hexagonal closed-packed (HCP) arrangement, with almost the ideal ratio of the $c$ and $a$ lattice parameters.\cite{2009_Moram_RepProgPhys72_036502} Thus we use an ideal HCP lattice of Ga sites in the KMC model. The use of the HCP lattice (P6$_3$/mmc symmetry) instead of the full wurtzite structure (P6$_3$mc symmetry) does not capture some features of the GaN structure, e.g. the asymmetry between ${[0001]}$ and ${[000\overline{1}]}$ due to polarity.

\begin{figure}
     \includegraphics[width=\columnwidth]{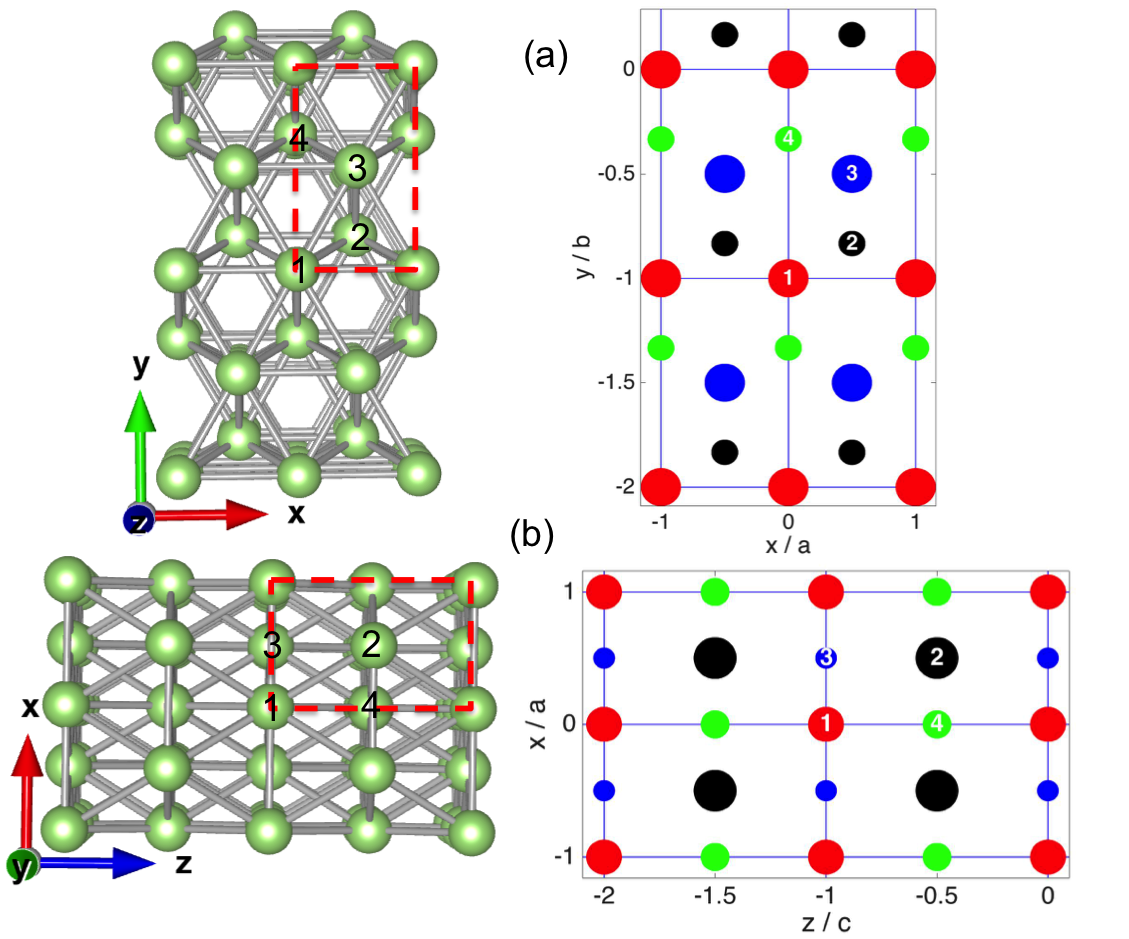}   
    \caption{Geometry of the Ga sites in GaN. (a) plan view of c-plane surface, cross-section of m-plane surface. (b) plan of m-plane, section of c-plane. Outlines show orthohexagonal unit cells.} 
    \label{fig:planes}
\end{figure}

An HCP lattice with lattice parameter $a$ can be described as an orthorhombic lattice using ``orthohexagonal'' coordinates,\cite{1965_Otte_PhysStatSol9_441} with four sites per unit cell and lattice parameters $a$, $b = \sqrt{3} a$, and $c = \sqrt{8/3} a$. Fig.~\ref{fig:planes} shows the geometry of the Ga sites on the c-plane and m-plane GaN surfaces. The three faces of the orthohexagonal unit cell correspond to the $(2 \overline{1} \overline{1} 0)$ a-plane, $(0 1 \overline{1} 0 )$ m-plane, and $(0 0 0 1)$ c-plane surfaces, which are normal to $a$, $b$, and $c$ in the $x$, $y$, and $z$ directions, respectively. 

As shown in cross section in Fig.~\ref{fig:planes}(b), for the c-plane surface, sites 1 and 3 form a layer at fractional coordinate $z/c = 0$ in each unit cell, while sites 2 and 4 form a layer at $z/c = 1/2$. We thus consider each of these layers to comprise a single monolayer (ML), and define the thickness of 1 ML to be $c/2$ for the c-plane. Note that in some of the literature, \cite{1999_Xie_PRL82_2749} this definition of monolayer is denoted as ``bilayer'', because of the nitrogen site associated with each Ga site. For the m-plane surface, with cross section shown in Fig.~\ref{fig:planes}(a), sites 1 and 2 have similar heights $y/b = 0$ and 1/6, while sites 3 and 4 have similar heights $y/b = 1/2$ and 2/3. We likewise consider each of these layers to comprise a single ML for the m-plane, with a thickness of $b/2$. 

\begin{table*}
\caption{ \label{tab:edgeE1} Step edge energies and Ehrlich-Schwoebel step edge barriers for c- and m-plane surfaces.}
\begin{ruledtabular}
\begin{tabular}{c|c|c|c|c|c|c} 
Surface & Step edge & Step & Step energy per & Step energy per  & Step edge diff. & Step edge diff.\\
orient. & normal & struct. & unit cell length  & unit length  & barrier (NN)  & barrier (NNN) \\ 
 & & & $(E_0)$ & $(E_0/a)$ & $(E_0)$ & $(E_0)$ \\ \hline 
c-plane & $y$ & A & 2 & 2.00 & $\infty$ & 0\\
c-plane & $y$ & B & 2 & 2.00 & 2 & 0\\
m-plane & $z$ & - & 1 & 1.00 & 4 & 2\\
m-plane & $x$ & - & 3 & 1.84 & $\infty$ & 0\\
\end{tabular}
\end{ruledtabular}
Note: Step edge diffusion barrier values in table show $\Delta E$ contribution only, not including $E_{barr}$ used in all diffusion jumps. 
\end{table*}

\subsection{Site and step energies}

In our KMC model, an energy $E_i$ is associated with each occupied Ga site $i$ that is a function of the number $N_i$ of occupied nearest-neighbor sites, which is $N_i = 12$ for sites in the bulk HCP lattice. The total energy of the system can be obtained by summing over the occupied sites
\begin{equation}
E_{tot} = \sum_{i ~ \text{occ.}} E_i,
\end{equation}
In general we use a simple linear function $E_i = - N_i E_0$, where each occupied nearest-neighbor site contributes an equal energy $-E_0$. The energy change that occurs if an atom is removed from or added a site corresponds to $\pm 2 E_i$; the factor of 2 accounts for the changes in the $N_i$ of the nearest-neighbor sites. This corresponds to an Ising model with nearest-neighbor interactions $J = -2E_0$.\cite{Giesen2001} This formula for $E_i$, based simply on counting of nearest neighbors, has also been used in some of the previous KMC simulations of crystal growth.\cite{Wang1994,Fu2008,Zaluska_Kotur2010} 

Values for the excess energies of non-ideal surface structural arrangements such as steps can be obtained from the bond-counting energy model described above. We have evaluated the structures and energies of various straight steps on c- and m-plane surfaces.\cite{Supplemental} Table~\ref{tab:edgeE1} summarizes the step edge energies for the lowest-energy steps. There are two low-energy step structures on the c-plane that have equal energies in this model. They are similar to the ``A'' and ``B'' steps on a $(1 1 1)$ surface of a face-centered cubic crystal.\cite{Giesen2001}

\subsection{Events and rates}

The evolution of the system occurs through two types of processes: diffusion jumps of atoms from occupied sites to unoccupied sites, and deposition of atoms into unoccupied sites from outside the system. A third potential process, evaporation of atoms from occupied sites to outside the system, is not considered. The neglect of evaporation is consistent with several other KMC simulations of GaN growth.\cite{Chugh2015, Zaluska_Kotur2010, Zaluska_Kotur2011, Zaluska_Kotur2012} The initial state of the simulation, representing a planar crystal surface at low temperature, has all sites occupied at locations below the surface plane, and all sites unoccupied above the plane. Periodic boundary conditions are applied to the simulation boundaries on the sides perpendicular to the surface plane. Prior to the start of growth, the surface was equilibrated at the growth temperature with zero deposition to establish the equilibrium vacancy and adatom concentrations. 

The rate of a diffusion jump of an atom from an initial occupied site $i$ to the unoccupied site $j$ is given by an Arrhenius expression based on transition state theory with an activation energy $E^A_{ij}$ that depends on the energy change $\Delta E \equiv 2 (E_j - E_i)$ due to the jump, and a barrier energy $E_{barr}$ representing the additional energy of the saddle point configuration during the jump. Supplemental Fig.~S1 shows a schematic of these energies. \cite{Supplemental} The average transition rate for the jump from site $i$ to site $j$ is given by
\begin{equation}
\Gamma_{ij} = \nu_0 \exp \left ( \frac{-E^A_{ij}}{kT} \right ),
\label{eq:Gamma}
\end{equation}
where $\nu_0$ is the attempt rate, $k$ is the Boltzmann constant and $T$ is the temperature. For ``uphill'' or ``downhill'' jumps, the activation energies are, respectively, 
\begin{eqnarray}
E^A_{ij} = \Delta E + E_{barr} &=& 2 (E_j - E_i) + E_{barr} ~ {\rm for} ~ \Delta E > 0, \nonumber \\
E^A_{ij} &=& E_{barr} ~~ {\rm for} ~~ \Delta E < 0. 
\end{eqnarray}
In general we use a single value of $E_{barr}$ for all diffusion jumps, independent of the initial and final states. This equally influences all jump rates through the same factor,
\begin{equation}
\Gamma_0 \equiv \nu_0 \exp(-E_{barr}/kT). 
\label{eq:Gamma0}
\end{equation}
So, changing the value of $E_{barr}$ just renormalizes the time scale of all diffusion processes in a temperature-dependent manner.

Our KMC model contains three scaling parameters, $E_0$, $\nu_0$, and $E_{barr}$, that can be adjusted to correspond to a given material. The value of $E_0$ sets the temperature scale, through the characteristic temperature $T_0 \equiv E_0 / k$. The value of $\nu_0$ sets the time scale in the high temperature limit, and the value of $E_{barr}$ sets the temperature dependence of the time scale.

For simplicity, simulation calculations are carried out in reduced energy and time units, where $E_0 = 1$ energy unit, $\nu_0 = 1$ (ut)$^{-1}$, where ``ut'' is the unit of time in the simulations, and we arbitrarily choose $E_{barr} = 0.3 E_0$. A reduced temperature $T / T_0$ is used. These reduced energy, temperature, and time units for the simulations can be related to actual units in experiments using known values of material properties, as described in Sec.~\ref{sec:exp}. 

To model crystal growth, we assume that deposition occurs at a defined rate into unoccupied sites at the crystal surface.\cite{Supplemental} The relative rates of deposition and surface diffusion govern the growth mode of the crystal.\cite{Tsao1993} Since the rate of surface diffusion and the equilibrium structure of the surface are temperature dependent, the growth mode varies as a function of temperature and deposition rate.  

\subsection{Two diffusion models: NN vs. NNN}\label{sec:model}

Two models for diffusion are considered. In the first model (``NN''), the neighbors of a site to which diffusion jumps can occur include only the 12 nearest neighbor sites. In the second model (``NNN''), under certain circumstances, diffusion jumps can also occur to some next-nearest-neighbor sites. An atom at site $i$ can jump to a vacant next-nearest-neighbor site $k$ if there are two intermediate sites $j_1$ and $j_2$ that are both nearest neighbors of $i$, $k$, and each other, and one is vacant and the other is occupied.\cite{2009_Plimpton_SAND2009-6226,SPPARKS_08jul2015} This provides an alternative model of the anisotropy of adatom diffusion on m-plane surfaces, and of the unusual type of jumps that can occur in the vicinity of step edges on surfaces, that affect the Ehrlich-Schwoebel (ES) barrier for transport across step edges from above.\cite{ehrlich1966atomic,schwoebel1966step} 

We have analytically evaluated the diffusion coefficients of adatoms on the c- and m-plane surfaces based on these two diffusion models.\cite{Supplemental} For isolated adatoms on the c-plane surface, diffusion is isotropic, with $D_{xx} = D_{yy} = \frac{3}{2} \Gamma_0 a^2$, and there is no difference between the NN and NNN models, because no next-nearest-neighbor sites fulfill the criterion in the NNN model.

\begin{figure}
     \includegraphics[width=0.40\textwidth]{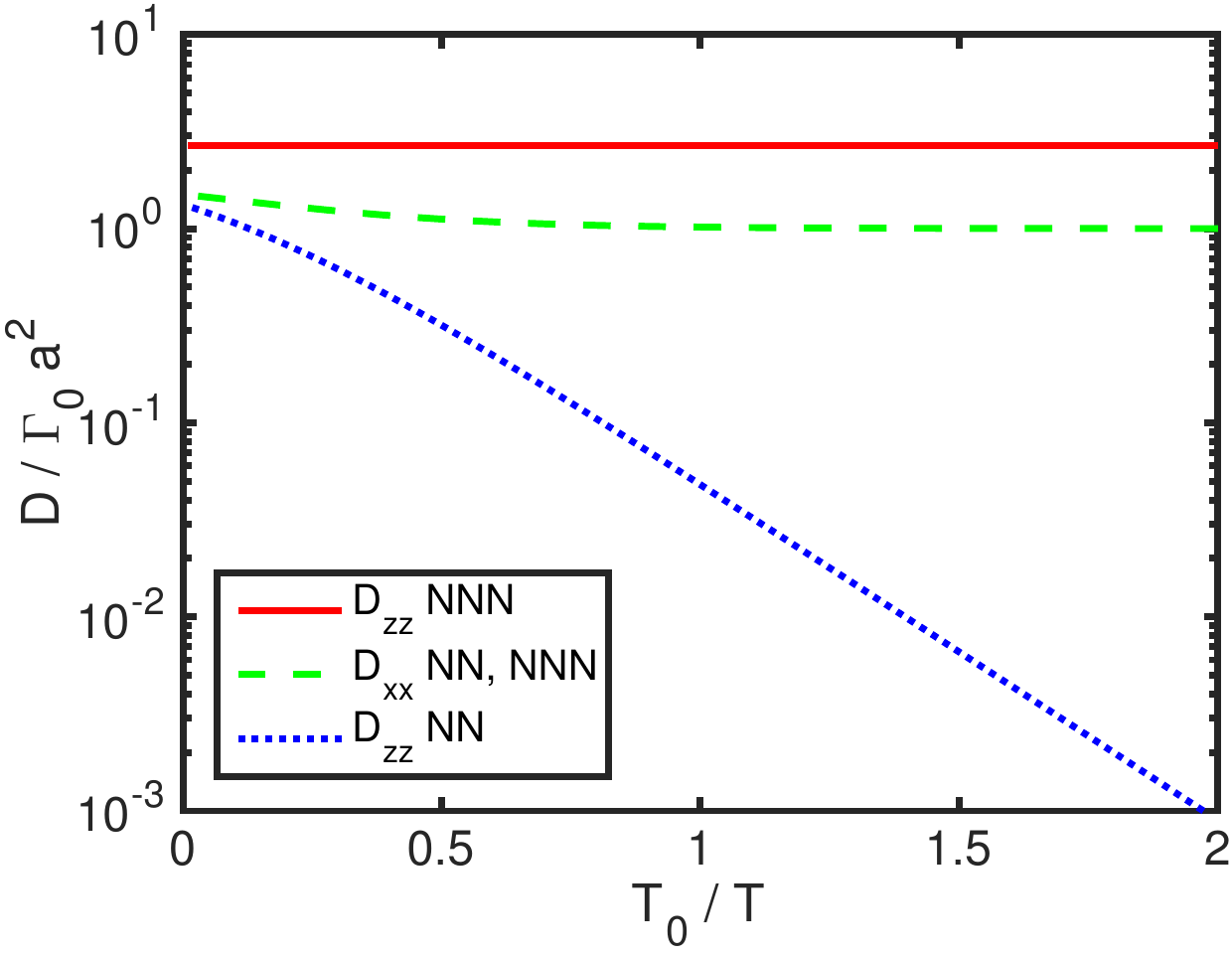}
    \caption{Predicted adatom diffusion coefficients $D_{xx}$ and $D_{zz}$ scaled by $\Gamma_0 a^2$, vs. inverse temperature $T_0 / T$ for the m-plane GaN surface. $D_{zz}$ depends upon the choice of NN or NNN diffusion model.}
    \label{fig:aniso}
\end{figure}

Adatom diffusion on the m-plane surface is more complex, because the surface has low symmetry with two different types of adatom sites, having different numbers of occupied nearest neighbors $N_i$ (and thus different energies and equilibrium occupancies). Fig.~\ref{fig:aniso} shows the values of the diffusion coefficients as a function of inverse temperature for the m-plane surface, for both the NN and NNN models. In the NN model, diffusion is strongly anisotropic at lower $T$, because the rows of low-energy sites running in the $x$ direction are separated by rows of high-energy sites, which increase the barrier to diffusion in the $z$ direction by $4 E_0$. However, in the NNN model, adatoms in the low-energy sites can jump directly to the next-nearest-neighbor low-energy row, bypassing the high-energy sites. While $D_{xx}$ is the same in both models, the anisotropy $D_{zz}/D_{xx}$ is inverted for NNN compared with NN. 

While we do not include an explicit ES step-edge barrier in our KMC model, as employed in other studies,\cite{Zaluska_Kotur2012,Zaluska_Kotur2011,Krzyzewski2016a} an effective barrier to diffusion down a step nevertheless arises because of the bonding geometry at the step.  Detailed examination of the geometry of neighbors at different step edges\cite{Supplemental} indicates that the NNN diffusion model produces a significantly smaller effective ES step-edge barrier, compared with the NN model. The effective ES barriers for the lowest-energy steps are summarized in Table~\ref{tab:edgeE1} for the two diffusion models. Because we have suppressed diffusion to sites with $N_i =1$, as described below, the ES barrier is infinite for two cases in the NN model. (If jumps to $N = 1$ sites had been allowed, the ES barriers would still have been large, $4 E_0$ or $6 E_0$, so their neglect has little effect on the growth simulation results.\cite{Supplemental}) In contrast, the NNN model has zero ES barrier in most cases (that is, no additional barrier above the standard barrier $E_{barr}$ for diffusion). For diffusion on the m-plane surface across a step normal to the $z$ direction, the NN model gives the same relatively high barrier of $4 E_0$  at the step edge as for adatoms diffusing in the $z$ direction on the terrace. The NNN model reduces the ES barrier to $2 E_0$. The NN and NNN models provide two extreme cases to demonstrate the effects of high and low ES barriers.   

\subsection{Implementation in SPPARKS}

We carried out the KMC simulations on a 3-dimensional lattice using the Stochastic Parallel PARticle Kinetic Simulator (SPPARKS) computer code.\cite{2009_Plimpton_SAND2009-6226,SPPARKS_08jul2015} The dynamics were calculated using the variable time step method known as the Gillespie or BKL algorithm.\cite{gillespie1977exact, Bortz1975} Details of the SPPARKS implementation are given in the supplemental material.\cite{Supplemental}

Some exceptions to the general rules stated above are made to control unwanted behavior of the simulation. To suppress diffusion of atoms or dimers from the crystal surface into the volume of unoccupied sites above the crystal, the energy of atoms with $N_i = 0$ or $1$ nearest neighbors is set to a large value. This has an additional effect of increasing the ES barrier for diffusion across certain step edges in the NN model, as described above.\cite{Supplemental} To suppress diffusion of single vacancies in the bulk, the diffusion barrier for jumps from sites with $N_i = 11$ to $N_j = 11$ is set to a high value. If this is not done, at higher temperatures, vacancies from the surface will diffuse through the crystal and accumulate at the lower boundary of the simulation. Since no periodic boundary conditions are applied at the upper and lower boundaries of the simulation, the sites there have fewer neighboring sites and are thus energetically favorable for vacancies. 

\begin{figure*}
       \includegraphics[width=0.8\textwidth]{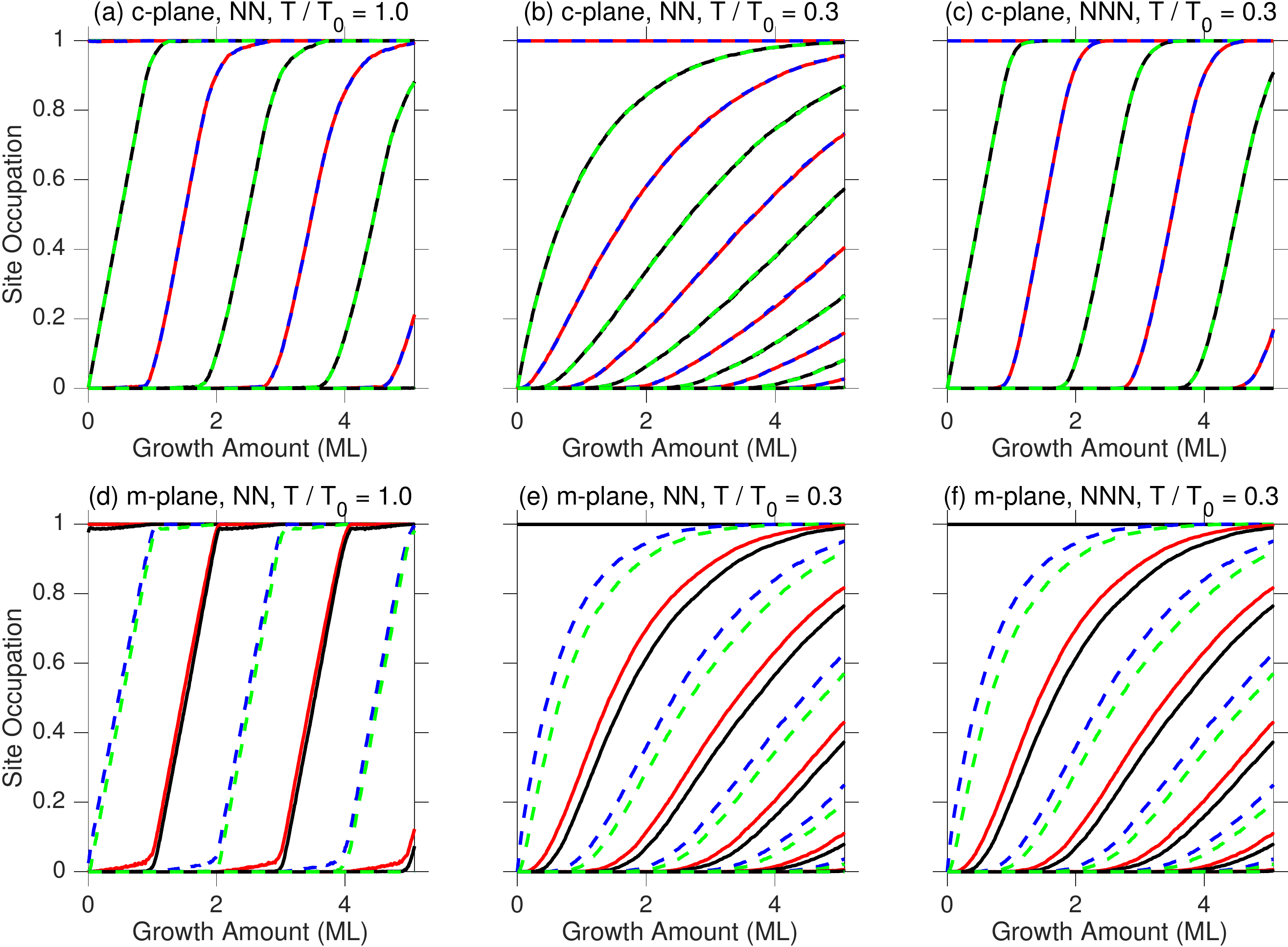}
     \caption{Fraction of sites occupied in each layer of unit cells during typical growth simulations. Growth rates were $G = 5.3 \times 10^{-6}$ or $5.0 \times 10^{-6}$ ML/ut for c- or m-plane, respectively. Colors for sites 1-4 are red, black, blue and green respectively.} 
     \label{fig:growth_coverage}
\end{figure*}
 
\section{Simulation Results}

\begin{figure*}
     \centering
          \includegraphics[width=\textwidth]{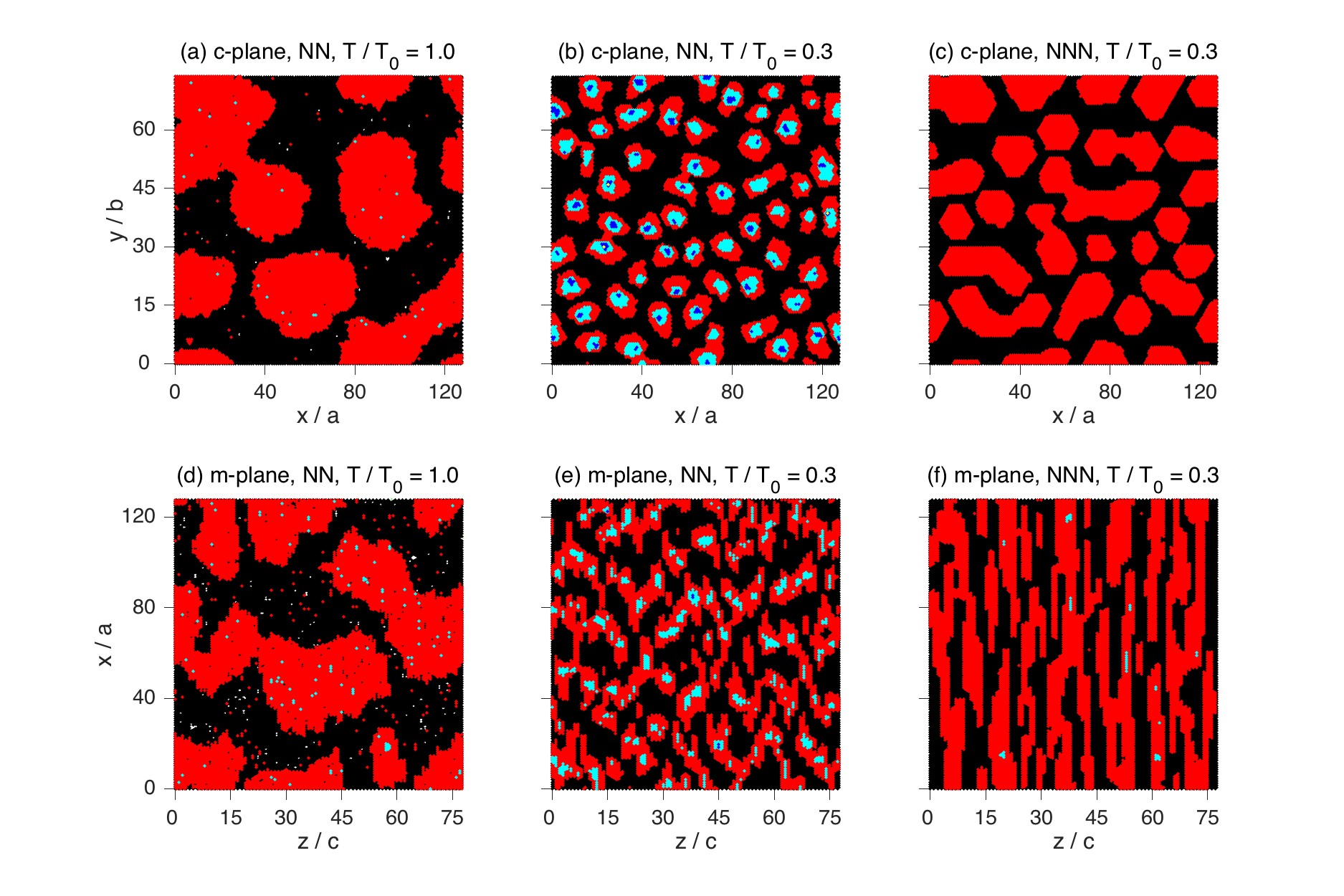}
     \caption{Surface islands after growth of 0.5 ML, for conditions of Fig.~\ref{fig:growth_coverage}.  The white, black, red, cyan, blue, and yellow colors represent the atoms at height of -1, 0, 1, 2, 3 and 4 ML above the initial surface.}  
     \label{fig:05ML_real}
\end{figure*}

\subsection{Layer-by-layer and 3D growth modes}

We investigated growth as a function of temperature $T$ and growth rate $G$, and observed the transition in the homoepitaxial growth mode\cite{Tsao1993}  between layer-by-layer (LBL) and 3-dimensional (3D) for the c- and m-plane surfaces, using both NN and NNN diffusion models. Fig.~\ref{fig:growth_coverage} shows examples of the occupation fraction of each of the four sites  in each layer of orthohexagonal unit cells shown in Fig.~\ref{fig:planes}, as a function of time (plotted as growth amount).

On the c-plane, one can see that sites in the same monolayer fill at the same rate under all conditions. At high temperature, Fig.~\ref{fig:growth_coverage}(a), growth occurs in LBL mode: each monolayer fills almost completely before there is significant occupation of the next monolayer. At low temperature in the NN model, Fig.~\ref{fig:growth_coverage}(b), growth occurs in 3D mode: several layers fill simultaneously. At high temperature, the behavior of the NNN model (not shown) is very similar to that of the NN model. However, in the NNN model, LBL growth persists even at low temperature, Fig.~\ref{fig:growth_coverage}(c).

On the m-plane, Figs.~\ref{fig:growth_coverage}(d-f), all four sites fill at different rates. However, sites in the same monolayer, as defined above, begin to fill at about the same time and track fairly closely, especially at high temperature. The dependence of the growth mode on temperature and diffusion model is qualitatively similar to that seen on the c-plane; in the NN model, a transition from LBL to 3D growth is seen between $T = T_0$ and $T = 0.3 T_0$, while in the NNN model, LBL growth persists to lower temperature. The linearity of the occupancy curves in Fig.~\ref{fig:growth_coverage}(d), with sharp changes in slope at the initiation or completion of each monolayer, indicates a nearly ideal LBL growth mode.\cite{Supplemental}

\subsubsection{Growth behavior in real space}

Figure~\ref{fig:05ML_real} shows the morphology of the islands on the surface after growth of 0.5 ML. The surface orientations, diffusion models, and conditions are the same as those for Fig.~\ref{fig:growth_coverage}. As expected, the average island size and spacing is larger at higher $T$, Fig.~\ref{fig:05ML_real} (a) and (d), than at lower $T$, reflecting the higher ratio of surface diffusion rate to growth rate. One can also see a significant population of adatoms and surface vacancies at higher $T$. For the NN diffusion model at $T$, Fig.~\ref{fig:05ML_real} (b) and (e), multi-layer islands are forming even after only 0.5 ML of growth, indicating that atoms deposited onto islands experience a large effect of the ES step-edge barrier. However, the islands remain single-layer at lower $T$ for the NNN model, Fig.~\ref{fig:05ML_real} (c) and (f), consistent with the lower ES barriers of the NNN diffusion model given in Table~\ref{tab:edgeE1}. There is little difference between NN and NNN (not shown) at higher $T$ for both c- and m-plane, because any ES barrier is easier to overcome. While islands on the c-plane surface are equiaxed at all temperatures, those on the m-plane surface become elongated in the $x$ direction at lower $T$ for the NNN model. For both c-plane and m-plane with the NNN model, the island edges become faceted at lower $T$, reflecting the low-energy step edge directions given in Table~\ref{tab:edgeE1}.
 
\begin{figure}
    \includegraphics[width=\columnwidth, trim={2cm 0.2cm 6cm 0cm}, clip]{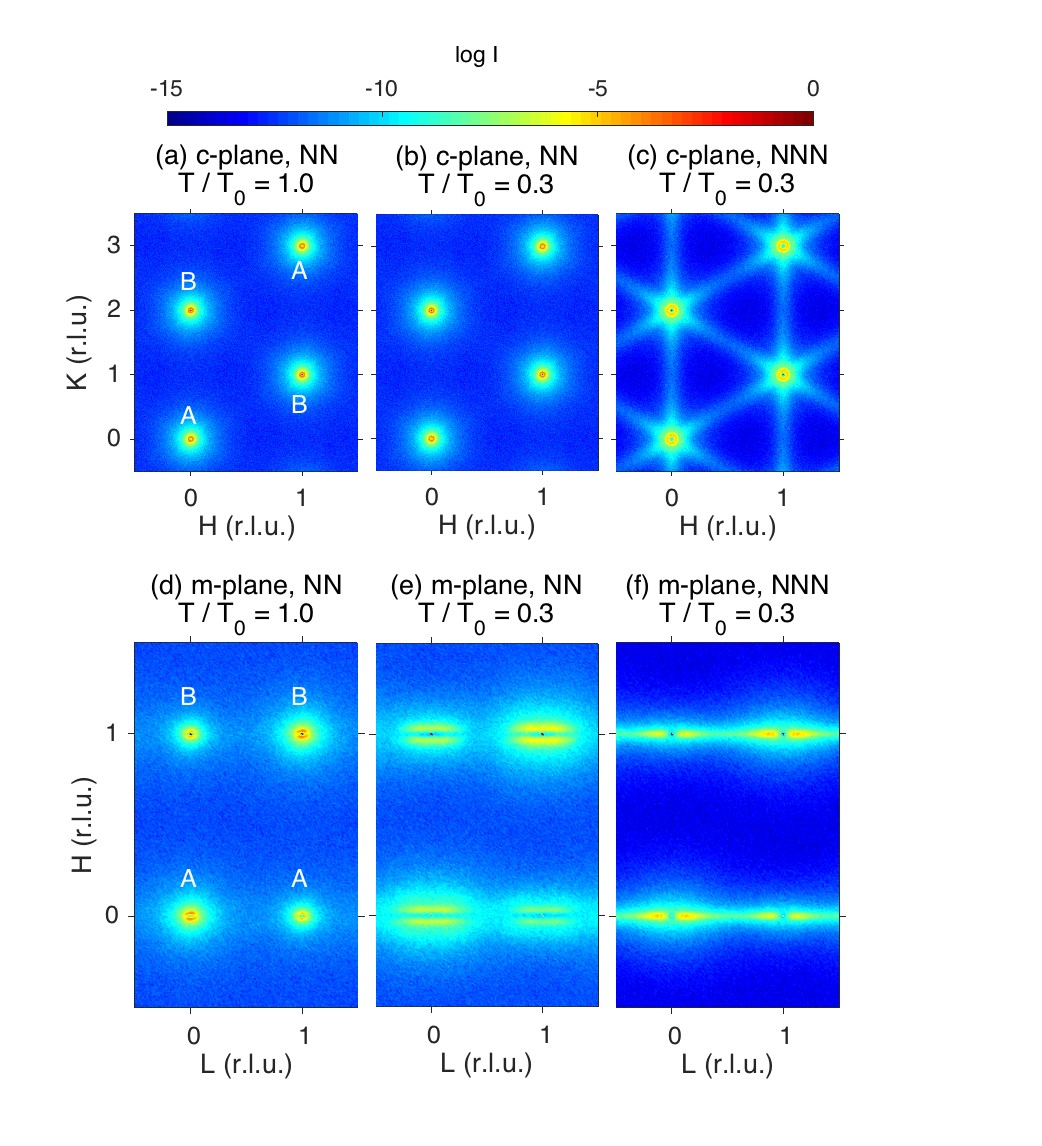}
     \caption{Reciprocal space intensity distributions after growth of $1/2$ ML, for conditions of Fig.~\ref{fig:growth_coverage}. For the c-plane surfaces (a) - (c), the distribution in the $HK$ plane is shown at $L = 1$. For the m-plane surfaces (d) - (f), the distribution in the $LH$ plane is shown at $K = 1$. The color scale represents the logarithm of the intensity. Labels B and A indicate Bragg and anti-Bragg peaks.}
     \label{fig:05ML_rcp}
\end{figure}

\begin{figure}
     \centering
    \includegraphics[width=\columnwidth, clip]{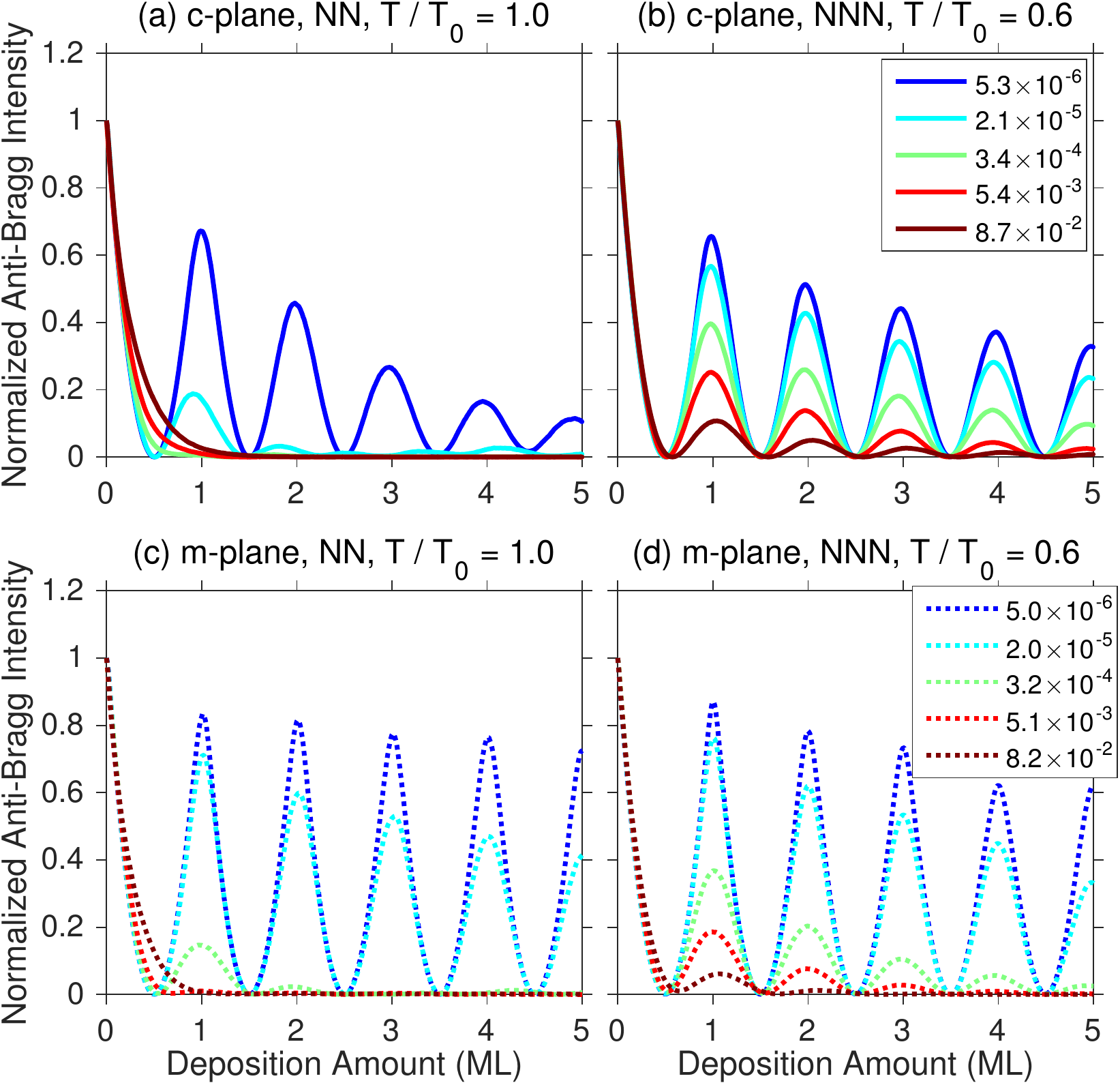}
     \caption{Anti-Bragg intensity at $(001)$ for c-plane and at $(010)$ for m-plane as a function of growth amount for growth rates shown in ML/ut, at fixed $T/T_0=1.0$ for NN kinetics and $T/T_0=0.6$ for NNN kinetics. Intensities are normalized to their initial values.} 
     \label{fig:I_Br_D}
\end{figure}

\subsubsection{Reciprocal space}\label{sec:rcp}

To quantitatively analyze the growth modes and island spacings, and to compare the results with surface X-ray scattering experiments, it is useful to calculate the intensity distribution in reciprocal space $I(H,K,L)$ (the square of the amplitude of the Fourier transform of the real space structure). Here we use reciprocal space coordinates $HKL$ of the orthohexagonal unit cells.\cite{1965_Otte_PhysStatSol9_441} Details are given in the Supplemental Material.\cite{Supplemental}

The scattered intensity is proportional to the square of the complex structure factor $F$, which we calculate as the sum of terms from the simulation volume and from a semi-infinite crystal substrate located beneath the simulation volume.
\begin{equation}
I(H,K,L) \propto \left | F_{sim} + F_{sub} \right |^2
\end{equation}
The substrate only contributes along the crystal truncation rods (CTRs)\cite{CTR} extending normal to the surface through the Bragg peaks at integer values of the in-plane reciprocal space coordinates ($H$ and $K$ for c-plane, $L$ and $H$ for m-plane).

The intensity distributions corresponding to typical 0.5 ML structures of Fig.~\ref{fig:05ML_real} are shown in Fig.~\ref{fig:05ML_rcp}. While results in Figs.~\ref{fig:growth_coverage} and \ref{fig:05ML_real} are for single simulations, results in Fig.~\ref{fig:05ML_rcp} and subsequent figures are obtained by averaging 16 simulations run with the same conditions but using different random number seeds. For the c-plane surfaces, Fig.~\ref{fig:05ML_rcp}(a-c), a slice through reciprocal space in the $HK$ plane at $L = 1$ is shown. Peaks appear where CTRs running in the $L$ direction cut through the plane shown. Labels show Bragg peak positions from the bulk crystal lattice, and ``anti-Bragg'' positions half-way between Bragg peaks along the CTRs. For m-plane surfaces, Fig.~\ref{fig:05ML_rcp}(d-f), a slice through reciprocal space in the $LH$ plane at $K = 1$ is shown. 

Diffuse scattering intensity around the CTR positions reflects the in-plane structure of the crystal surface. For c-plane, the distribution of equiaxed islands with correlated positions gives rings of intensity in reciprocal space. The radius of the ring is inversely proportional to the spacing of the islands. When the island edges become faceted at lower $T$, the intensity distribution in reciprocal space shows streaks normal to the facets. For m-plane, the increasing anisotropy at lower $T$ gives very different distributions in the $H$ and $L$ directions. For the NN model, there are well-defined satellite peaks split in the $H$ direction, while for the NNN model the satellites are split in the $L$ direction. The satellites split along $H$ indicate a well-defined island spacing along $x$, while satellites split along $L$ indicate a well-defined island spacing along $z$. The amount of splitting is inversely proportional to the island spacing.

\subsubsection{Evolution of anti-Bragg intensity}

Alternate monolayers scatter exactly out of phase at anti-Bragg positions, giving good sensitivity to surface morphology from islands. Fig.~\ref{fig:I_Br_D} shows the average intensity at the anti-Bragg position in reciprocal space ($HKL = 001$ for c-plane, $010$ for m-plane) as a function of growth amount at the different growth rates given in the legends, for each surface orientation and diffusion model. For the NN or NNN models, temperatures $T/T_0 = 1.0$ or $0.6$ are shown, respectively. For all cases, we see oscillations in intensity that reflect the growth mode. For ideal LBL growth, in which each monolayer completely coalesces before islands of the next monolayer form, the anti-Bragg intensity will make parabolic oscillations with equal maxima at integer monolayer amounts of deposition, and zero intensity at half-integer monolayer amounts. As the growth deviates from LBL and becomes 3D, new layers begin to form before the growth of the previous layer finishes, and the amplitude of the oscillation decreases and eventually disappears. In all cases in Fig.~\ref{fig:I_Br_D}, we see this decrease in oscillation amplitude as growth rate increases, indicating the transition from LBL to 3D growth mode. We note that the positions of the maxima also shift away from integer monolayer amounts of deposition in the transition to 3D growth, and that the shift is to lower amounts for the NN model and to higher amounts for the NNN model. This reflects a change in the nature of the multilayer height distribution as the ES barrier is varied.
 
\begin{figure}
     \centering
     \includegraphics[width=0.8\columnwidth]{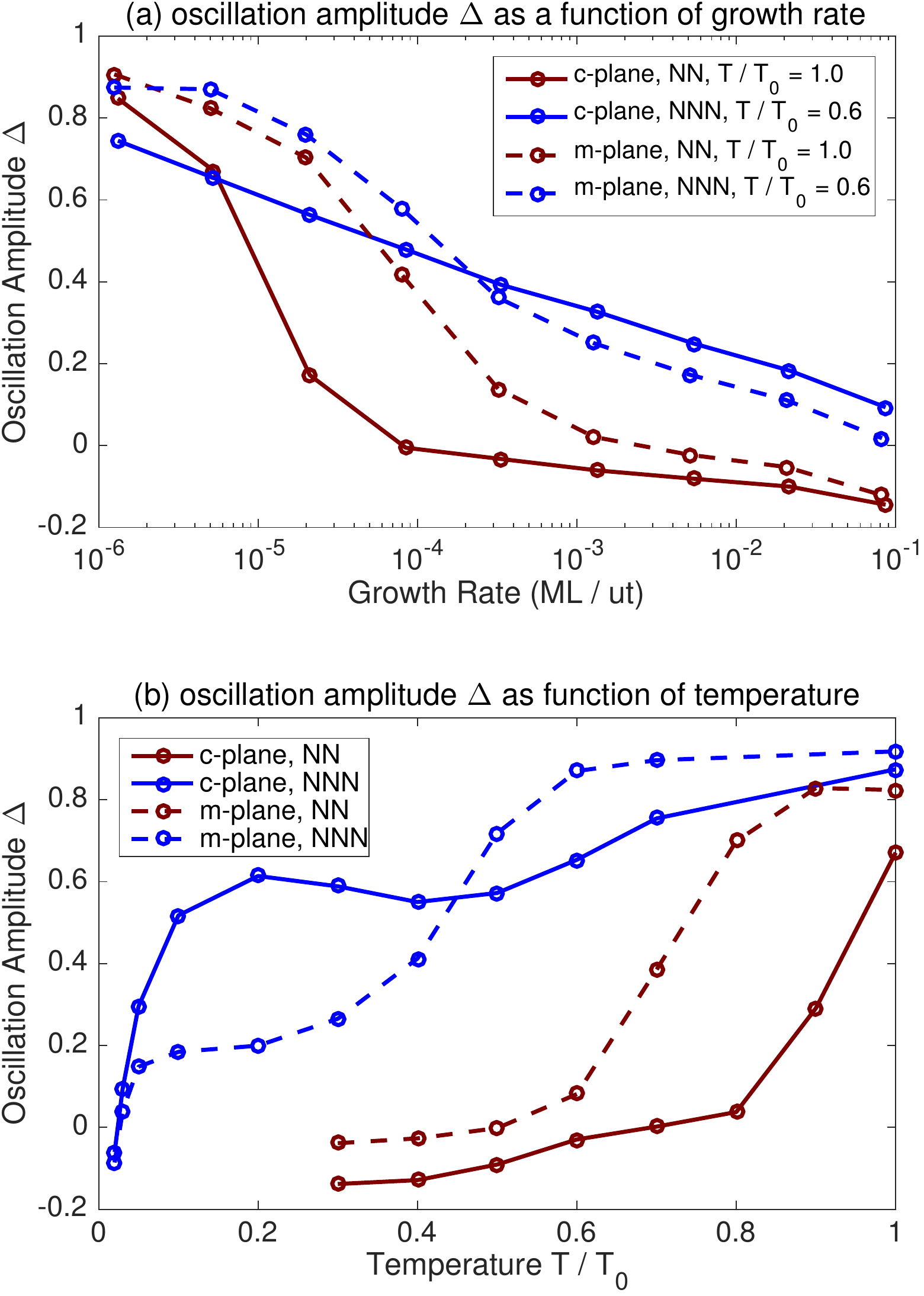}
     \caption{Comparison of oscillation amplitude $\Delta \equiv (I_{1}-I_{1/2})/I_{0}$ for c-plane and m-plane with NN and NNN kinetics. (a) $\Delta$ as a function of growth rate $G$ at fixed $T/T_0 = 1.0$ or $0.6$ for NN or NNN, respectively. (b) $\Delta$ as a function of $T/T_0$ at fixed $G=5.3$ or $5.0 \times 10^{-6}$ ML/ut for c-plane or m-plane, respectively.}  
     \label{fig:Delta}
\end{figure}

\begin{figure}
     \centering
     \includegraphics[width=\columnwidth]{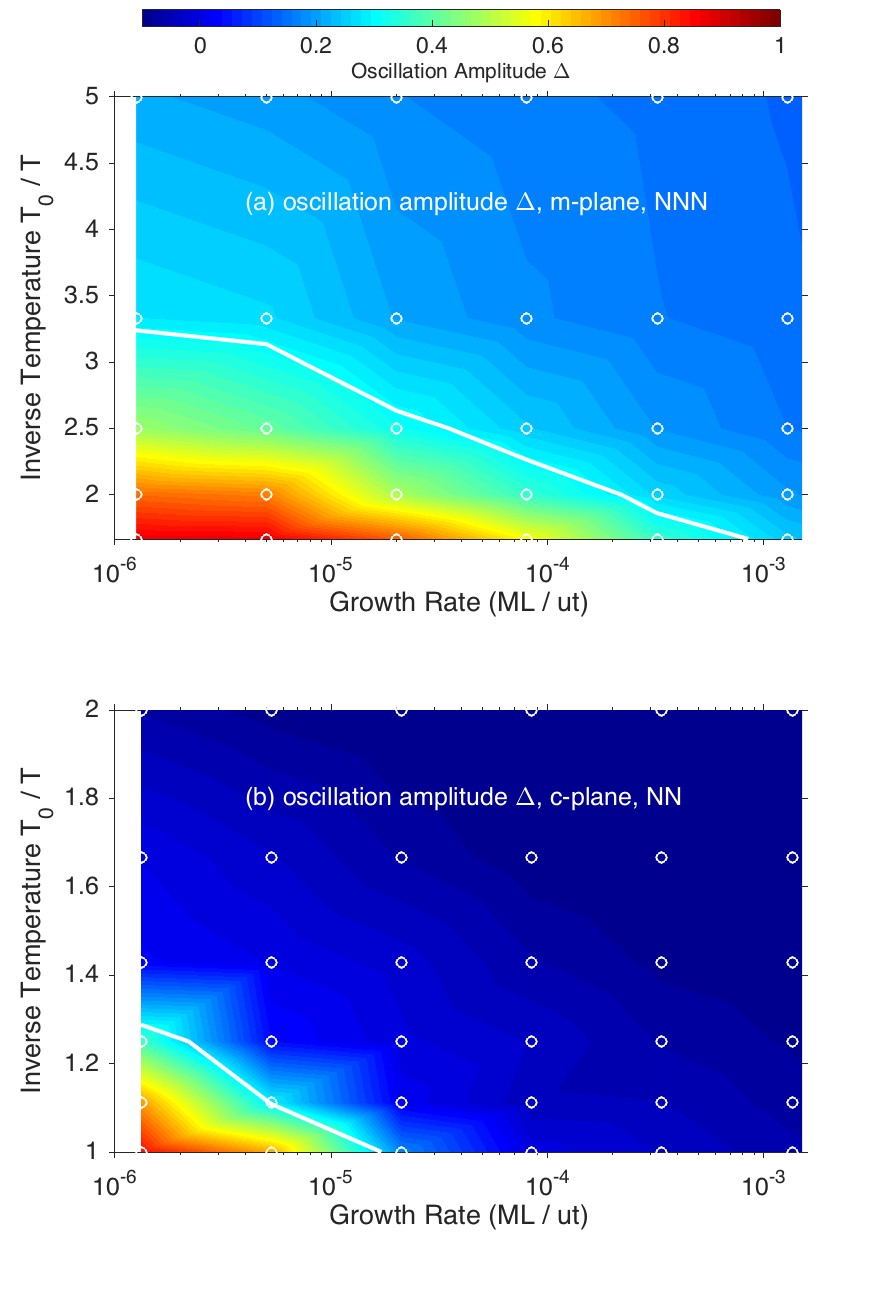}
        \caption{Plots of oscillation amplitude $\Delta$ as a function of inverse temperature and growth rate for (a) m-plane with NNN kinetics and (b) c-plane with NN kinetics. White circles show the conditions for the simulation data points, and the color scale gives the interpolated value of $\Delta$. The white contour at $\Delta = 0.3$ indicates the LBL-3D growth mode boundary in each case. Note that $T_0/T$ range shown differs in (a) and (b).}
        \label{fig:DT}
\end{figure}

\subsubsection{Growth mode transition between LBL and 3D}\label{sec:growthmode}

To characterize the transition between LBL and 3D growth, we define a normalized oscillation amplitude $\Delta \equiv (I_{1}-I_{1/2})/I_{0}$, where $I_{1}$, $I_{1/2}$, and $I_{0}$ are the anti-Bragg intensity for growth amounts of $1$, $1/2$ and $0$ ML, respectively.\cite{StephensonAPL1999,Perret2014} Values of $\Delta$ approaching unity indicate perfect LBL growth, while values approaching zero (or becoming negative) indicate 3D growth. Here we choose a value of $\Delta = 0.3$ to define the boundary between LBL and 3D growth.

Figure~\ref{fig:Delta}(a) shows the $\Delta$ values as a function of growth rate $G$ for c-plane (solid curves) and m-plane (dashed curves) with the NN model at a reduced temperature $T/T_0=1.0$ (red curves) or the NNN model at $T/T_0=0.6$ (blue curves). In all cases $\Delta$ decreases as growth rate increases, indicating a transition from LBL to 3D growth. Fig.~\ref{fig:Delta}(b) shows the $\Delta$ values as a function of temperature at fixed growth rates $G$ in ML/ut. Again the transition from LBL to 3D is reflected in the decrease in $\Delta$ as temperature decreases. For both m-plane and c-plane, the transition occurs at a lower $T$ with NNN kinetics compared with NN kinetics. The change is especially large for c-plane, compared with m-plane. For the NNN model on m-plane, the transition as a function of $T$ appears to occur in two steps, with some drop in $\Delta$ at intermediate $T$, followed by a sharp drop at low temperature, $T/T_0 < 0.1$. For the NNN model on c-plane, the primary drop occurs at low $T$. This behavior reflects two different mechanisms for the transition to 3D growth: adatoms become trapped on top of islands because of an ES step-edge barrier, or adatoms have insufficient time for surface diffusion compared with incoming deposition. The 3D growth mode will occur in all cases at low $T$ due the second mechanism. A transition to 3D can occur at intermediate $T$ due to the first mechanism, depending upon the magnitude of the ES barrier. For m-plane, where the NNN model reduces but does not eliminate the ES barrier for the dominant steps normal to the $z$ direction, we see the first mechanism is shifted to lower $T$ but not eliminated. For c-plane, the elimination of the ES barrier leaves only the second mechanism.

Figure~\ref{fig:DT} shows the behavior of $\Delta$ as a function of both growth rate and inverse temperature, for m-plane with NNN kinetics and c-plane with NN kinetics. (We focus on these two cases because they agree best with experiments, as described below. Results for the m-plane with NN kinetics were also obtained, and are shown in the supplemental material.\cite{Supplemental}) Contours of constant $\Delta$ on this plot are reasonably straight lines, indicating that the temperature dependence of the growth rate at the LBL-to-3D boundary $G_{3D}$ can be described by the Arrhenius expression
\begin{equation}
G_{3D} = A_{3D} \exp(-E_{3D}/kT).
\label{eq:3D}
\end{equation}
For each temperature, we interpolated between the $\Delta$ values to determine the boundary growth rate $G_{3D}$ giving $\Delta = 0.3$, and fit these values with Eq.~\ref{eq:3D}. Parameters obtained from these fits are given in Table \ref{tab:delta_fits}.

\begin{table}
\caption{ \label{tab:delta_fits} Parameters of Eq.~\ref{eq:3D} obtained from fits to the growth rate as a function of temperature at the LBL-3D boundary defined by $\Delta = 0.3$.}
\begin{ruledtabular}
\begin{tabular}{c|c|c|c} 
Surface & Diffusion & $E_{3D} / E_0$ & $\log_{10} A_{3D}$  \\ 
Orient. & Model & & (ML/ut) \\ \hline
c-plane & NN & $8.41 \pm 0.57$ & $-1.20 \pm 0.28$ \\
m-plane & NNN & $3.81 \pm 0.09$ & $-0.42 \pm 0.08$ \\
\end{tabular}
\end{ruledtabular}
\end{table}

\subsection{Island spacing and step-flow growth}

\subsubsection{Diffuse scattering}

Characteristics of the island structure on the crystal surface can be obtained by analyzing the in-plane diffuse scattering around the CTRs at 0.5 ML of growth, such as that shown in Fig.~\ref{fig:05ML_rcp}. 

\begin{figure}
     \centering
     \includegraphics[width=0.8\columnwidth]{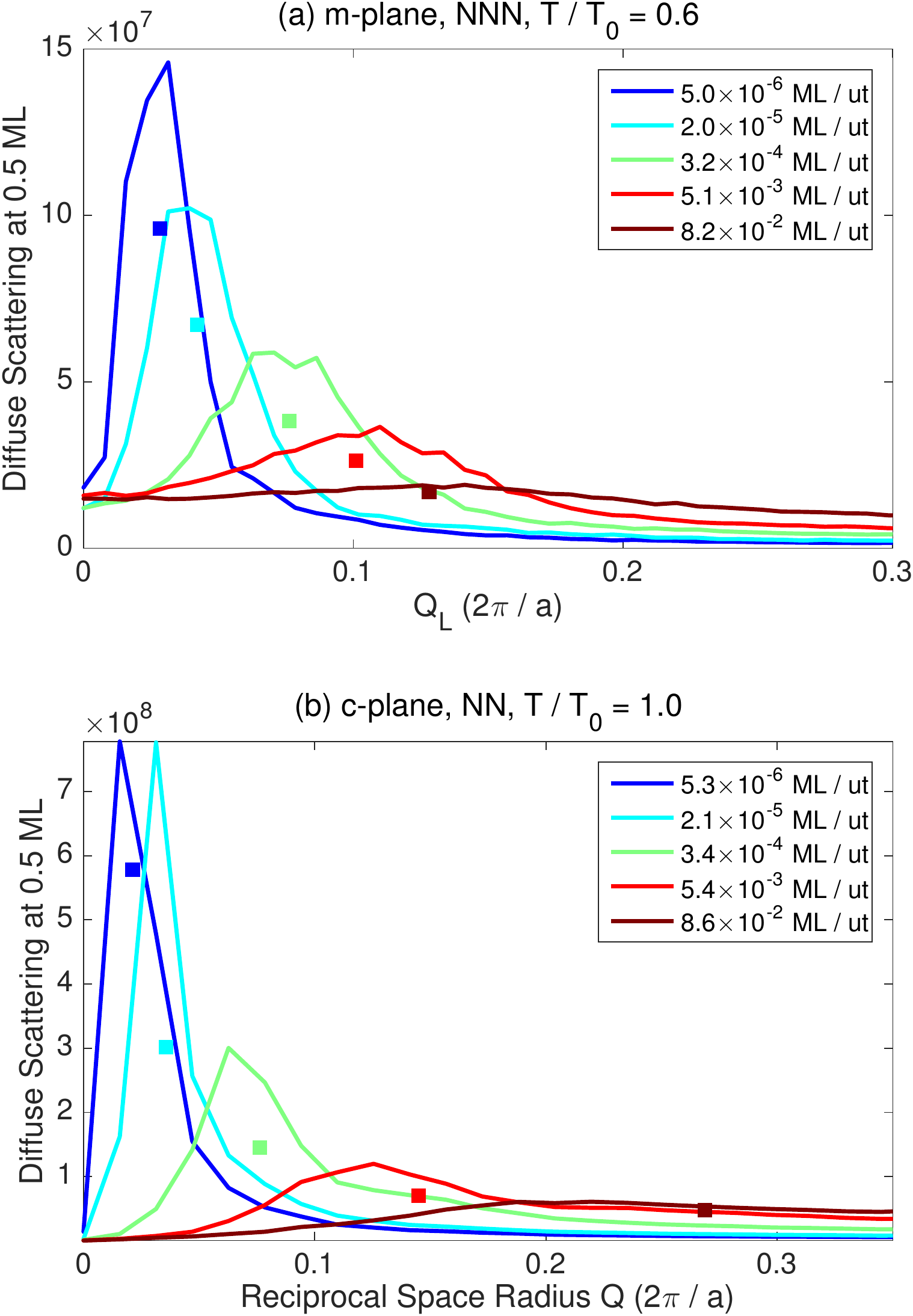}
        \caption{Diffuse scattering profiles near the CTR at 0.5 ML for different growth rates given in legend. (a) m-plane, $T/T_0 = 0.6$, NNN kinetics; (b) c-plane, $T/T_0 = 1.0$, NN kinetics. Square symbols show the extracted positions $\overline{Q}$.} 
        \label{fig:COM}
\end{figure}

For the m-plane surface, the diffuse scattering is not isotropic. We see symmetric satellite peaks on both sides of the CTR, displaced in the in-plane directions $L$ or $H$ for the NNN or NN results, respectively. For the NNN model, we analyzed the intensity distribution averaged over $H$ within a region centered on the CTR. Fig.~\ref{fig:COM}(a) shows typical plots of the diffuse intensity as a function of $Q_L$ around the $(010)$ CTR on the m-plane surface with NNN kinetics, for various growth rates at $T/T_0 = 0.6$. We extracted the positions by self-consistently calculating the center of mass $\overline{Q}$ of the intensity distribution for $Q_L > 0$ within the range from $0.4 \, \overline{Q}$ to $1.6 \, \overline{Q}$ using an iterative procedure. The extracted value of $\overline{Q}$ for each curve on Fig.~\ref{fig:COM} is shown by the square symbol.

For the c-plane surface, a nearly isotropic ring of intense diffuse scattering indicates an isotropic arrangement of islands with correlated spacings. To characterize these rings, we performed an azimuthal average centered on the CTR position to obtain the intensity as a function of reciprocal space radius $Q$ (in units of $2 \pi / a$). Fig.~\ref{fig:COM}(b) shows typical plots of the diffuse intensity around the $(001)$ CTR on the c-plane surface with NN kinetics, for various growth rates at $T/T_0 = 1.0$. As for the m-plane, we see peaked intensity distributions, the positions of which are inversely related to the average island spacings. 

\subsubsection{Island spacing}

\begin{figure}
     \includegraphics[width=\columnwidth]{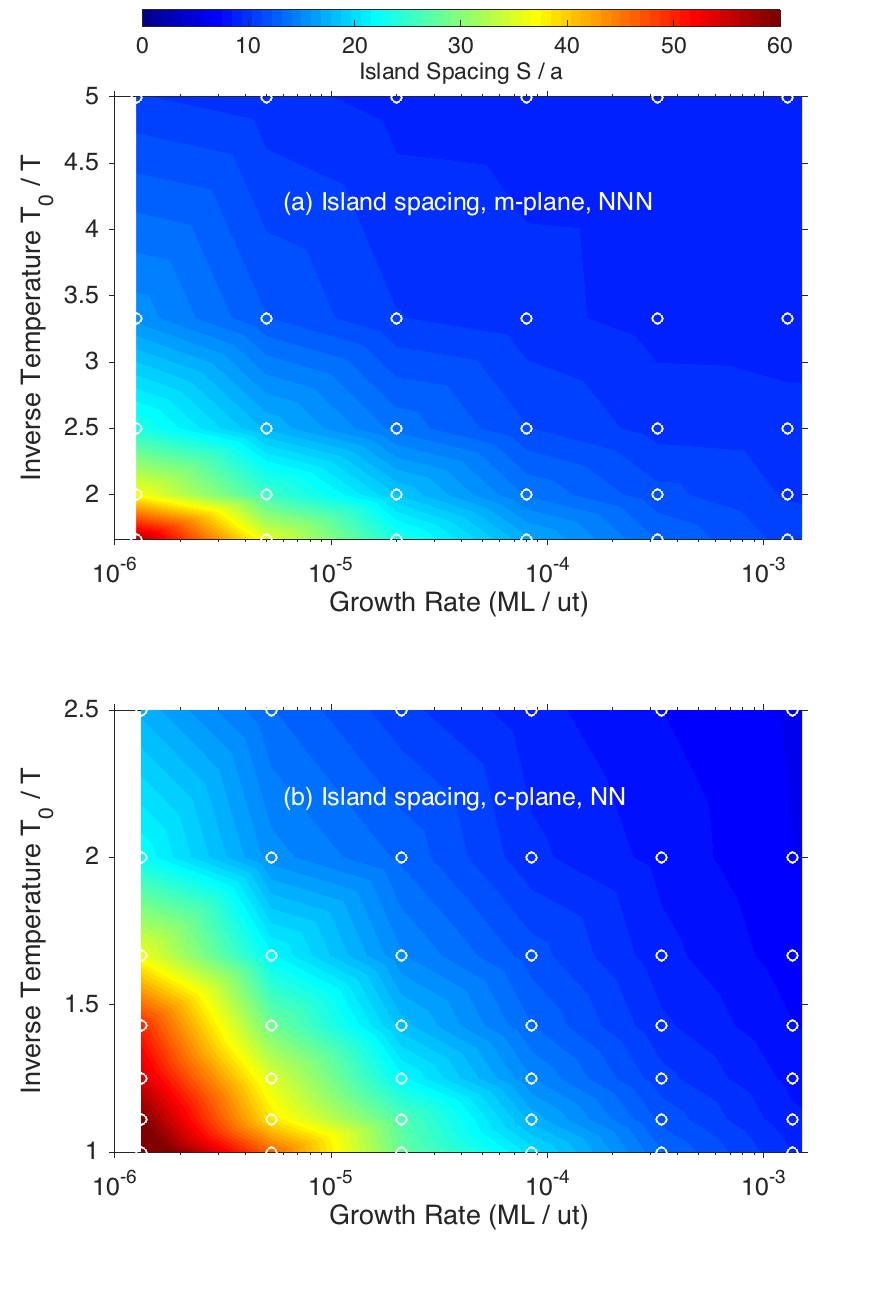}
     \caption{Island spacing $S/a$ at 0.5 ML as a function of inverse temperature and growth rate for (a) m-plane with NNN kinetics and (b) c-plane with NN kinetics. The white circles show the conditions for the simulation data points, and the color scale gives the interpolated value of $S/a$. Note that the $T_0/T$ range shown differs in (a) and (b).}
     \label{fig:ST}
\end{figure}

The average island spacing in real space $S$ can be obtained from the peak position in reciprocal space using $S = 2 \pi / \overline{Q}$. Fig. \ref{fig:ST} shows the island spacing at 0.5 ML as a function of growth rate and inverse temperature for m-plane with NNN kinetics and c-plane with NN kinetics. 

\begin{figure}
     \centering
     \includegraphics[width=0.8\columnwidth]{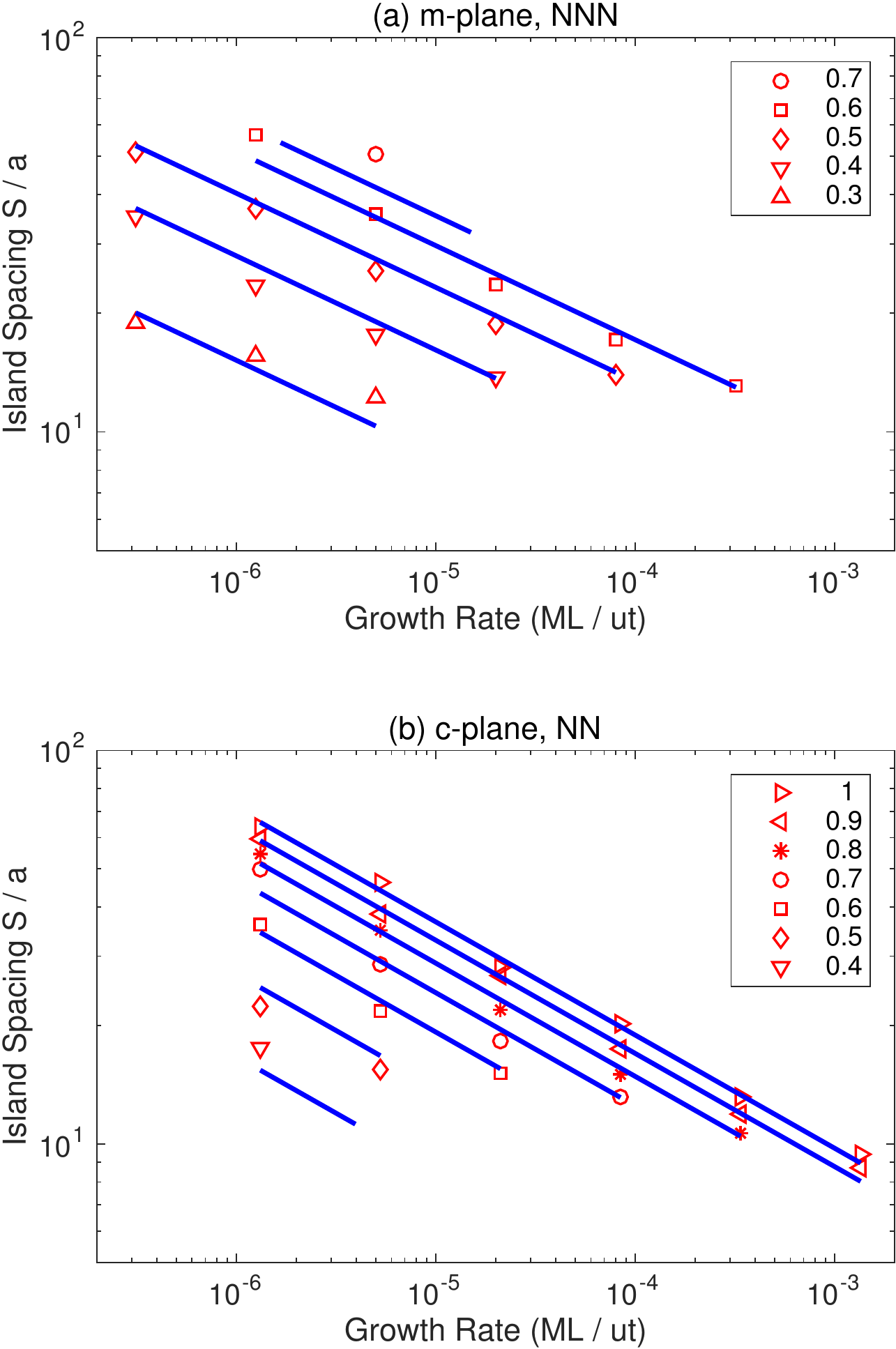}
     \caption{Island spacing $S/a$ at 0.5 ML as a function of growth rate at various $T/T_0$ given in legend, for (a) m-plane with NNN kinetics and (b) c-plane with NN kinetics. Lines are fit of all points in each plot to Eq.~\ref{eq:S}.} 
     \label{fig:SF}
\end{figure}

In the LBL region, the temperature and growth rate dependence of the island spacing can be modeled using an expression from nucleation theory,\cite{Evans1994}
\begin{equation}
S/a = (G/G_S)^{-n} \exp(-nE_S/kT),
\label{eq:S}
\end{equation}
where island spacing $S$ has been scaled by the $a$ lattice parameter.
We have fit the island spacings for higher temperatures and lower growth rates to this expression, as shown in Fig.~\ref{fig:SF}, and obtained the fit parameters, $n$, $E_s$ and $G_s$, given in Table~\ref{tab:spacing_fits}.

\begin{table}
\caption{ \label{tab:spacing_fits} Parameters of Eq.~\ref{eq:S} obtained from fits to island spacings.}
\begin{ruledtabular}
\begin{tabular}{c|c|c|c|c} 
Surface & Diffusion & $n$ & $E_S / E_0$ & $\log_{10} G_S$  \\ 
Orient. & Model &  &  & (ML/ut) \\ \hline
c-plane & NN & $0.288 \pm 0.007$ & $3.36 \pm 0.13$ & $1.90 \pm 0.10$ \\
m-plane & NNN & $0.239 \pm 0.014$ & $3.06 \pm 0.19$ & $3.39 \pm 0.29$ \\
\end{tabular}
\end{ruledtabular}
\end{table}

\subsubsection{Step-flow transition}

In this study we consider surfaces oriented exactly on the crystallographic c- or m-planes. On vicinal surfaces away from these exact orientations, long-range step arrays are present and growth can occur by the step-flow (SF) mode, in which adatoms attach to the step arrays in preference to nucleating islands. The boundary between SF and LBL growth modes is expected to occur when the island nucleation spacing $S$ equals the terrace width (step spacing) $W$ of the vicinal surface.\cite{Tsao1993} While we do not directly model vicinal surfaces in this study, an expression for the boundary can be obtained by substituting $S = W$ into Eq.~\ref{eq:S} and solving for the growth rate as a function of $T$ and $W$,
\begin{equation}
G_{SF} = G_S  \, (W/a)^{-1/n} \exp(-E_S/kT).
\label{eq:SF}
\end{equation}

\begin{figure}
\centering
     \includegraphics[width=0.8\columnwidth]{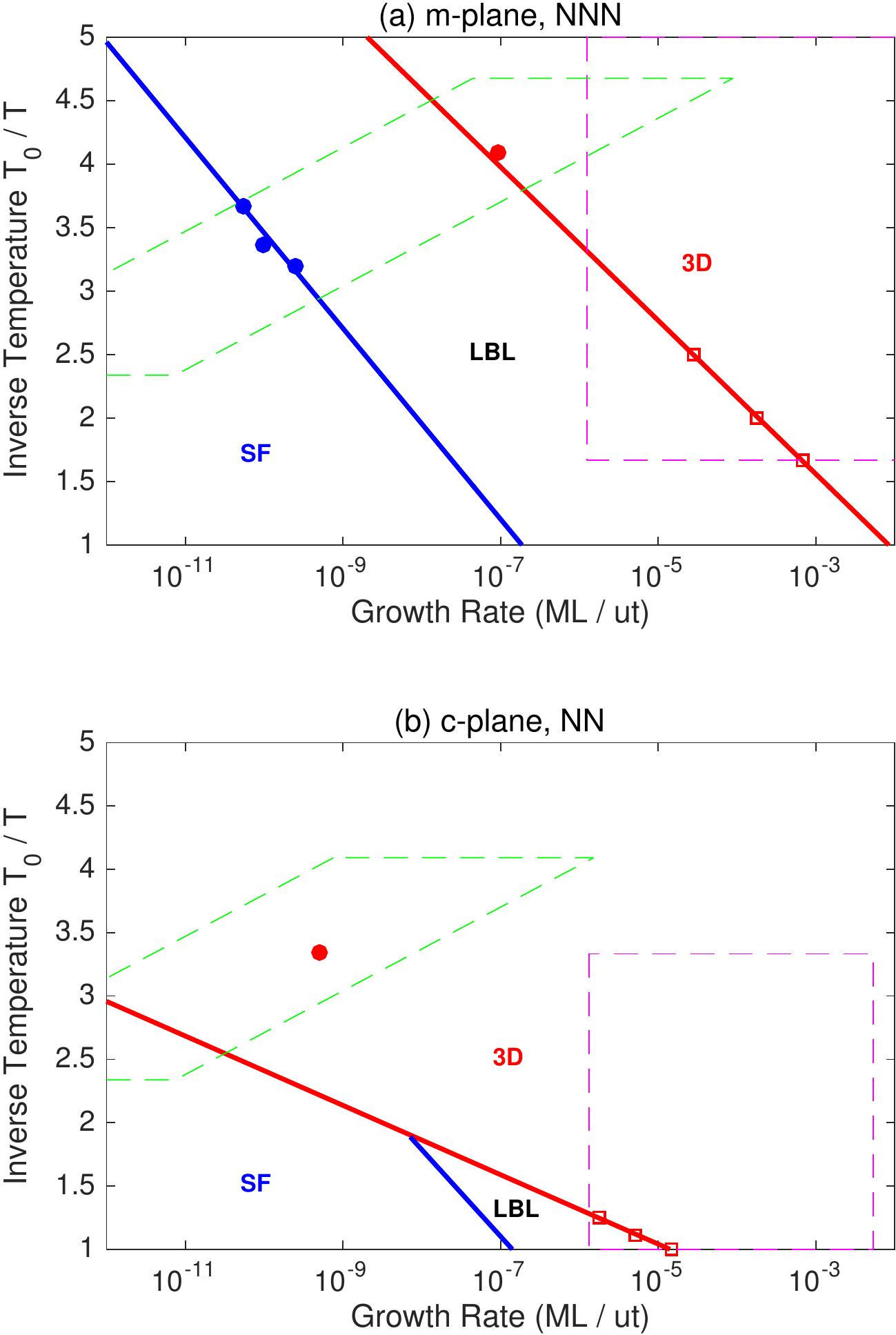}
     \caption{Predicted homoepitaxial growth mode transitions on GaN surfaces, for (a) m-plane with NNN kinetics and (b) c-plane with NN kinetics. Red and blue lines show the boundaries between 3-dimensional (3D), layer-by-layer (LBL) and step-flow (SF) growth modes, for a terrace width $W/a = 125$. Magenta and green dashed boundaries show ranges of conditions for simulations and experiments, respectively.} 
     \label{fig:GM}
\end{figure}

Figure~\ref{fig:GM} shows the predicted growth mode boundaries for the m-plane surface with NNN kinetics, and the c-plane surface with NN kinetics, as a function of growth rate $G$ and inverse temperature. The LBL-3D boundaries are obtained from the fits of the simulation values of $G_{3D}$ using Eq.~\ref{eq:3D} given in Table~\ref{tab:delta_fits}. The SF-LBL boundaries are from Eq.~\ref{eq:SF}  using the values in Table~\ref{tab:spacing_fits} and $W/a = 125$, which corresponds to the experimental results.\cite{Perret2014} Note that these boundaries are extrapolated outside of the region directly investigated in the simulations. The range of LBL growth is much wider for the m-plane than for the c-plane, primarily because the large ES barrier for the c-plane with NN kinetics expands the 3D growth region.  

\section{Comparison with Experiment}\label{sec:exp}

\subsection{Length, energy, temperature, and time units for GaN MOVPE}

To compare the KMC results quantitatively with experiment, we can relate the dimensionless units of the simulations to units appropriate for GaN growth by MOVPE. The length scale for GaN is given simply by its crystal lattice parameters, which are $a = 0.32$~nm and $c = 0.52$~nm at 1000~K.\cite{2009_Moram_RepProgPhys72_036502}

To estimate the energy scale for GaN, the total energy per lattice site, $12 E_0$, can be equated to the enthalpy of formation of solid GaN from vapor GaN, $\Delta H_f^{solid} - \Delta H_f^{vapor} = -3.38$ eV per molecule\cite{Przhevalskii1998} at 1000 K. This gives an energy unit for the simulation of $E_0 = 0.282$ eV, and a characteristic temperature of $T_0 = 3275$ K. Thus typical GaN MOVPE growth temperatures of 1000 to 1600~K correspond to simulation conditions in the range $T/T_0 = 0.3$ to $0.5$.

The time unit $t_0$ (s/ut) is a temperature dependent quantity that depends on the $\nu_0$ and $E_{barr}$ values that enter into the surface diffusion coefficients, e.g. $D = \frac{3}{2} a^2 \nu_0 \exp(-E_{barr}/kT)$ for the c-plane surface. An expression for $t_0$ can be obtained by taking the ratio of $\Gamma_0$ in simulation units (ut$^{-1}$) and in experimental units (s$^{-1}$),
\begin{equation}
t_0 = \frac{\exp(- 0.3 T_0 / T)}{\nu_0 \exp (-E_{barr} / kT )}.
\label{eq:t0}
\end{equation}
Note that the time unit $t_0(T)$ is a function of temperature. Thus the growth rate $G/t_0$ in ML/s (experiment units) changes with temperature at fixed growth rate $G$ in ML/ut (simulation units). Values for $t_0$ are estimated below based on the observed SF-LBL growth mode boundary for m-plane GaN MOVPE.

\subsection{Island height and shape}

The heights of the islands observed during LBL growth in the simulations (e.g. Figs.~\ref{fig:growth_coverage} and \ref{fig:05ML_real}) were $c/2$ for the c-plane and $b/2$ for the m-plane, which we defined to be 1 ML in each case. These heights agree with those observed in experiments for both c-plane\cite{StephensonAPL1999,ThompsonJECS, MunkholmAPL} and m-plane.\cite{Perret2014}

\begin{figure}
     \centering
     \includegraphics[width=0.70\columnwidth]{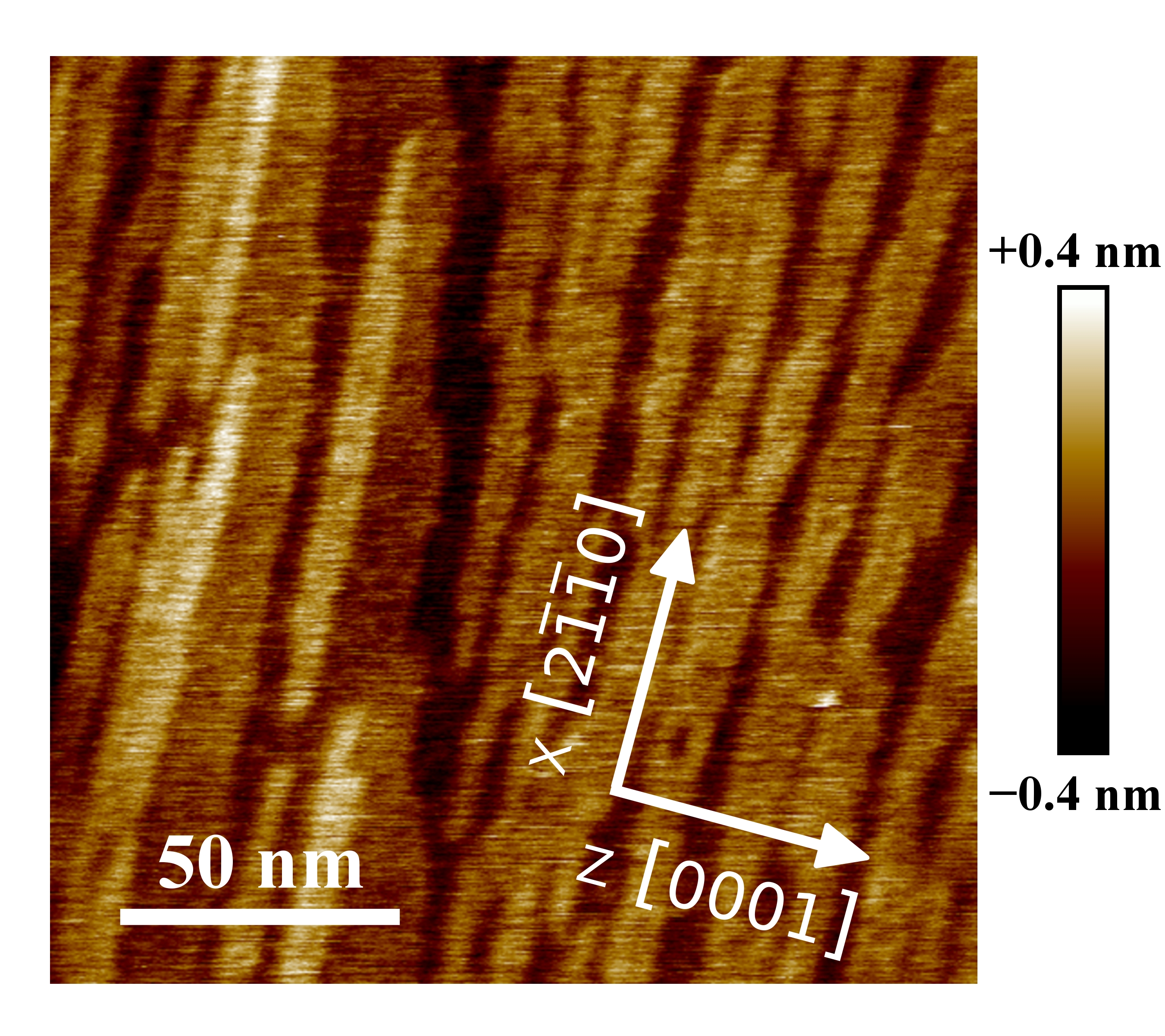}
        \caption{Atomic force microscopy (AFM) topography image of m-plane surface with 0.5 ML coverage of islands grown under LBL conditions.\cite{Perret2014,Perret_unpub}. Islands are elongated normal to $z$, as in Fig.~\ref{fig:05ML_real}(f) for the NNN model.} 
        \label{fig:AFM}
\end{figure}

The arrangement and anisotropy of the island shapes for m-plane seen in the simulations differ significantly between the NN and NNN kinetic models at lower $T$, as shown in Figs. \ref{fig:05ML_real}(e) and (f). In particular, the NNN model gives islands that are elongated in the $x$ direction and have a well-defined spacing in the $z$ direction. This is in agreement with experimental observations of island shapes on m-plane surfaces, as shown in Fig.~\ref{fig:AFM}. 

\begin{table}
\caption{ \label{tab:scales} Reduced temperature and time units, and calculated adatom diffusion coefficients for c- amd m-plane surfaces, as a function of actual temperature, using estimated parameters for GaN MOVPE: $T_0 = 3275$ K, $E_{barr} = 2.05$ eV, and $\nu_0 = 3\times10^{18}$ s$^{-1}$. Values of $D_{zz}$ for m-plane use the NNN model.} 
\begin{ruledtabular}
\begin{tabular}{c|c|c|c|c|c} 
$T$ & $T/T_0$ & $t_0$ & $D$ (c) & $D_{xx}$ (m) & $D_{zz}$ (m)\\
(K)  &       & (s) & (cm$^2$/s) & (cm$^2$/s) & (cm$^2$/s) \\ \hline
1000 & 0.31 & $2.5 \times 10^{-9}$ & $2.3 \times 10^{-7}$ & $1.5 \times 10^{-7}$ & $4.0 \times 10^{-7}$\\
1250 & 0.38 & $2.6 \times 10^{-11}$ & $2.6 \times 10^{-5}$ & $1.8 \times 10^{-5}$ & $4.7 \times 10^{-5}$\\
1500 & 0.46 & $1.3 \times 10^{-12}$ & $6.3 \times 10^{-4}$ & $4.2 \times 10^{-4}$ & $1.1 \times 10^{-3}$\\
\end{tabular}
\end{ruledtabular}
\end{table}

\subsection{Growth mode boundaries}

As shown in Fig.~\ref{fig:GM}, the simulations predict a much larger region of LBL growth for m-plane than c-plane surfaces. This result is consistent with experiments,\cite{ThompsonJECS,Perret2014} in which a direct transition between SF and 3D growth modes is typically observed on c-plane, while a large intervening region of LBL growth is observed on m-plane.

The observed growth rate at the SF-LBL boundary on m-plane GaN\cite{Perret2014} can be described by an Arrhenius expression
\begin{equation}
G_{SF}^{obs} = A_{SF}^{obs} \exp(-E_{SF}^{obs}/kT),
\end{equation}
with parameters $E_{SF}^{obs} = 2.83 \pm 0.27$~eV and $\log_{10}[A_{SF}^{obs}$~(ML/s)$] = 13.1 \pm 1.4$. By equating $G_{SF}^{obs}$ in ML/s with $G_{SF}/t_0$ from Eqs.~\ref{eq:SF} and \ref{eq:t0}, we can obtain expressions for the parameters that give the time scale for GaN MOVPE,
\begin{equation}
E_{barr} = E_{SF}^{obs} - E_S + 0.3 E_0.
\end{equation}
\begin{equation}
\log(\nu_0) = \log(A_{SF}^{obs}) -\log(G_S) + n^{-1}\log(W/a),
\end{equation}
Using the terrace width $W = 40 \pm 10$~nm corresponding to the experiments\cite{Perret2014} and the simulation value of $n = 0.239 \pm 0.014$, this gives $E_{barr} = 2.05 \pm 0.28$ eV and $\log_{10}[\nu_0$~(s$^{-1})] = 18.5 \pm 1.8$. Values of $t_0$ estimated using these parameters are given in Table~\ref{tab:scales}. Also shown are the calculated surface diffusion coefficients.\cite{Supplemental}

Using the correspondence between experimental and simulation temperature and time scales given in Table~\ref{tab:scales}, the regions of temperature and growth rate investigated in recent experiments\cite{Perret2014} are shown in Fig.~\ref{fig:GM}. The boundaries at fixed experimental growth rates of 0.001 and 2 ML/s appear as diagonal lines when plotted in simulation units of ML/ut, because of the temperature dependence of $t_0$. Except at low temperature, the region investigated directly in the current simulations corresponds to much higher growth rates than studied in experiments. The three experimental values of $G_{SF}^{obs}$ used to obtain the correspondence are likewise plotted in simulation units (blue dots) on Fig.~\ref{fig:GM}(a).

The experimentally observed value of $G_{3D}$ at the LBL-3D transition for m-plane\cite{Perret2014} is plotted (red dot) in simulation units on Fig.~\ref{fig:GM}(a). It lies very close to the boundary obtained from the simulation results (red line). This represents a quantitative agreement of the relative positions of the SF-LBL and LBL-3D boundaries between the experiments and the simulations using the NNN model for the m-plane surface.

Experiments on c-plane surfaces\cite{Perret2014} show a direct transition between SF and 3D growth modes, with no intervening LBL mode. This qualitatively agrees with the prediction in Fig.~\ref{fig:GM}(b) for the region of experimental conditions. However, the experimentally observed value for $G_{3D}$ (red dot) is significantly higher than the boundary predicted by simulations using the NN model for the c-plane surface (red line). This quantitative difference may indicate that the effective ES barriers produced by the NN model are too large to accurately represent the c-plane, and that behavior intermediate between those of the NN and NNN simulations would agree better with experiment. However, because we have used results from m-plane to determine the correspondence between experimental and simulation time scales, other effects may also contribute to the difference observed for c-plane.

\section{Discussion and Conclusions}

The simulations presented here of growth on the c- and m-plane surfaces of a hexagonal crystal provide both agreement with, and insight into, experimental results for MOVPE growth of GaN. We see different behavior on the two surfaces because of the different bonding configurations. The simulations employ two different models for diffusion kinetics: NN, which allows diffusion jumps only between nearest-neighbor sites; and NNN, which also allows some next-nearest-neighbor jumps. These produce higher and lower effective Ehrlich-Schwoebel step-edge barriers, respectively, and have a significant impact on the crystal growth modes. We have mapped the transitions as a function of temperature $T$ and growth rate $G$ among the three homoepitaxial growth modes: three-dimensional (3D); layer-by-layer (LBL); and step-flow (SF). We have also determined the dependence on $T$ and $G$ of the island spacing $S$ that develops during LBL and 3D growth. Quantitative comparison of these results to experiment allows us to estimate underlying fundamental quantities, such as the surface diffusion coefficients given in Table~\ref{tab:scales}.

On both the c- and m-plane surfaces, the island heights found in the simulations during LBL growth are one half of the orthohexagonal unit cell dimension ($c/2$ and $b/2$, respectively). This is in agreement with experimental studies of layer-by-layer growth in GaN MOVPE,\cite{StephensonAPL1999,ThompsonJECS, MunkholmAPL,Perret2014} indicating that the basic nearest-neighbor bond counting energetics of the KMC model is generally applicable to this system. The simulations show that both the typical island shapes at lower temperatures and the growth mode transitions differ between c- and m-plane, and between NN and NNN kinetics. Overall, the NN model with high ES barrier provides better agreement with experiment for the c-plane, while the NNN model with low ES barrier provides better agreement for m-plane. In particular, the very narrow region of LBL growth found for c-plane with NN kinetics, and the typical shape of islands on m-plane formed under NNN kinetics (elongated perpendicular to $[0001]$ and highly correlated parallel to $[0001]$), are both in agreement with X-ray\cite{Perret2014} and AFM measurements. This indicates that the elongated islands observed on m-plane have their origin primarily in anisotropic equilibrium step-edge energies rather than in adatom diffusion kinetics, since the latter is almost isotropic in the NNN model. Similar conclusions were reached in studies of islands on dimerized Si (001) \cite{Clarke199191} and Ag (110) \cite{1997_Ferrando_PRB56_R4406}.

Our study indicates that the direct transition from SF to 3D growth observed for growth on GaN c-plane surfaces can be attributed to a high ES barrier due to the bonding arrangement at steps on a close-packed surface. Previous KMC studies on c-plane growth also emphasized the effect of the ES barrier on island nucleation, finding that smoother films are obtained when the ES barrier is screened.\cite{Kaufmann2016} This agrees with our finding that the LBL range is wider on c-plane when using the NNN model. 

Our simulations found that the island spacing $S$ under LBL growth conditions obeys a negative power-law dependence on growth rate, Eq.~\ref{eq:S}, with exponent $n = 0.24$ for m-plane and $n = 0.29$ for c-plane. Recent experimental results \cite{Perret_unpub} for MOVPE on m-plane GaN are in agreement with this value. Growth of anisotropic islands in submonolayer epitaxy has been previously considered in KMC modeling. \cite{2001_Heyn_PRB63_033403,Clarke199191,Evans2006} The power law for island density as a function of growth rate shows an exponent smaller than that for the isotropic case, with $n = 1/4$ for a critical nucleus size equal to one. However, we observe a similar low exponent in our simulations on c-plane surfaces, where the island shapes are isotropic.

Several topics for future work can be identified. There is considerable interest in GaN growth on surface orientations in addition to c- and m-plane,\cite{2015_Kelchner_JCrystGrowth411_56} and it is straightforward to extend the KMC model and methods developed here to other orientations. While we have estimated the boundary between LBL and SF growth based on island spacing, it will be of interest to explicitly consider growth on vicinal surfaces with steps, which can be done by using helical boundary conditions.\cite{Krzyzewski2016} For simplicity we have neglected evaporation from the surface in this work. Inclusion of evaporation will provide a second surface transport mechanism qualitatively different from surface diffusion, that undoubtedly becomes important at higher temperatures.\cite{Mitchell2001} Since the KMC model gives an exact arrangement as a function of time for the atomic positions, it will be very valuable in predicting the results of coherent X-ray experiments (such as X-ray photon correlation spectroscopy \cite{Shpyrko2014} or coherent diffraction imaging \cite{2013_Abbey_JOM65_1183}) that are becoming feasible. The model also can be extended to account for full the wurtzite structure of GaN, with two atom types and additional parameters derived from experiment or ab initio theory. This will allow study of the effects of polarity. Finally, the NN and NNN models employed here give fairly extreme values for high and low effective ES barriers. The best agreement with experiments may be obtained from a model between these two extremes. 

\section{Acknowledgements}
We gratefully acknowledge support provided by the Department of Energy, Office of Science, Basic Energy Sciences, Scientific User Facilities (KMC model development) and Materials Sciences and Engineering (reciprocal space analysis and experiments), and computing resources provided on Blues and Fusion, high-performance computing clusters operated by the Laboratory Computing Resource Center at Argonne National Laboratory.

%

\clearpage


\section*{Supplementary material}

# Supplemental material for "Kinetic Monte Carlo simulations of GaN homoepitaxy on c- and m-plane surfaces"

Dongwei Xu,* Peter Zapol, and G. Brian Stephenson

*Materials Science Division, Argonne National Laboratory, Argonne, IL 60439*

Carol Thompson

*Department of Physics, Northern Illinois University, DeKalb IL 60115*
(Dated: revision 5 presub: November 8, 2016)

Here we provide details of the assumptions used, and supplemental results of, the kinetic Monte Carlo model developed for GaN homoepitaxy. These include details of sites and energy states used, determination of step energies and Ehrlich-Schwoebel step-edge barriers for c-plane and m-plane surfaces, and development of analytical expressions for adatom diffusion on c-, m-, and a-plane surfaces. We provide details of the implementation in SPPARKS and information regarding the typical number of events and computer time involved in the simulations. Finally we provide analysis of equilibrium layer occupations, formulas for calculating the intensity in reciprocal space observed in x-ray scattering experiments, and results on m-plane surfaces using the nearest-neighbor (NN) diffusion model.

## I. SITES AND ENERGIES

Table S.1 lists the positions of the four sites in each orthohexagonal unit cell. Note that since 5 monolayers (ML), each 1/2 unit cell thick, were typically used as the starting condition for simulations of the c- and m-plane surfaces, the initial surface was terminated with a half-completed unit cell (as shown in Fig. 1 of the main paper).

The energy associated with each occupied site $i$ is given by a simple linear function of the number $N_i$ of occupied nearest-neighbor sites, $E_i = -N_i E_0$. This is essentially a counting of nearest-neighbor bonds, where each bond has an energy $-2E_0$. The factor of two accounts for the two sites associated with each bond. The exceptions to this formula are for occupied sites with $N_i = 0$ or $N_i = 1$, which are assigned a very high energy (essentially infinite) in order to suppress "evaporation" of atoms or dimers into the sites above the crystal surface. Since adatoms with $N_i = 1$ can occur at certain step edges (see below), we checked the effect of suppressing their occupancy by performing some simulations in which the $N_i = 1$ sites are assigned energy $E_i = -E_0$ according to the linear formula. There was no significant effect on the parameters extracted from the growth behavior (e.g. island sizes or $\Delta$ values).

Figure S.1 shows the relationship between the energy

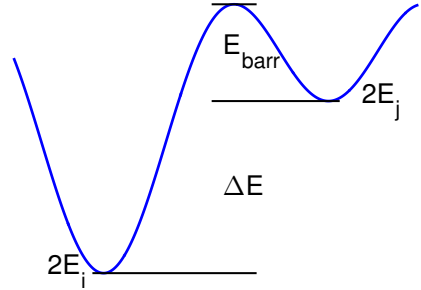

FIG. S.1. Energy diagram showing definition of $E_{barr}$ and energy change for a diffusion jump $\Delta E \equiv 2(E_j - E_i)$ relative to the energies $E_i$ of occupied sites $i$.

change $\Delta E \equiv 2(E_j - E_i)$ due to a diffusion jump from site $i$ to site $j$, and the diffusion barrier $E_{barr}$ defined relative to the site with higher energy.

## II. STEP STRUCTURES AND ENERGIES

Here we analyze how step structure affects the step edge energies and the effective Ehrlich-Schwoebel (ES) barriers for various high-symmetry steps, on the c- and m-plane surfaces within our KMC model. The step edge energies and ES barriers obtained below are summarized in Table S.2.

### A. c-plane

For the c-plane surface, we analyze step edges normal to the $\hat{y}$ and $\hat{x}$ directions.

TABLE S.1. Fractional coordinates of the Ga sites in one orthohexagonal unit cell, and colors used in Figs. 1 and 4 of the main paper.

| Site # | Color | $x/a$ | $y/b$ | $z/c$ |
|--------|-------|-------|-------|-------|
| 1 | red | 0 | 0 | 0 |
| 2 | black | 1/2 | 1/6 | 1/2 |
| 3 | blue | 1/2 | 1/2 | 0 |
| 4 | green | 0 | 2/3 | 1/2 |

TABLE S.2.  Step edge energies and Ehrlich-Schwoebel barriers for c- and m-plane surfaces.

| Surface orient. | Step edge normal | Step struct. | Figure # (energy) | Step energy per unit cell length ($E_0$) | Step energy per unit length ($E_0/a$) | Figure # (ES barr.) | Step edge diff. barrier (NN) ($E_0$) | Step edge diff. barrier (NNN) ($E_0$) |
|---|---|---|---|---|---|---|---|---|
| c-plane | $\hat{y}$ | A | S.2 | 2 | 2.00 | S.5, S.6 | $\infty$ | 0 |
| c-plane | $\hat{y}$ | B | S.3 | 2 | 2.00 | S.5, S.6 | 2 | 0 |
| c-plane | $\hat{x}$ | - | S.4 | 4 | 2.31 | S.7 | 2 | 0 |
| m-plane | $\hat{z}$ | A | S.8 | 1 | 1.00 | S.14 | 4* | 2 |
| m-plane | $\hat{z}$ | B | S.9 | 3 | 3.00 | S.15 | $\infty$ | 4 |
| m-plane | $\hat{x}$ | A | S.10, S.12 | 3 | 1.84 | S.16, S.18 | $\infty$ | 0 |
| m-plane | $\hat{x}$ | B | S.11 | 5 | 3.06 | S.17 | 2 | 0 |
| m-plane | $\hat{x} - \hat{z}$ | - | S.13 | 4 | 2.09 | S.19 | 2 | 0 |

Notes: Step edge diffusion barrier values in table show $\Delta E$ contribution only, not including $E_{barr}$ used in all diffusion jumps. Values for low-energy steps, highlighted in cyan, are shown in Table I of the main paper. *This ES barrier is the same as the barrier for NN diffusion in the $\hat{z}$ direction on the terrace.

TABLE S.3.  Analysis of step energy for triangular islands with A- or B-step edges on the c-plane surface, as illustrated in Figs. S.2 or S.3 for $s = 5$ atoms. Formulas for number $n_N$ of terrace and island atoms having excess energy, as well as excess energy per atom $\Delta E_N$ and total excess energy $\Delta E_{tot}$, as a function of occupied nearest-neighbor number $N_i$ and island edge length $s$.

| atom type | $N_i$ | $n_N$ A steps | $n_N$ B steps | $\Delta E_N$ ($E_0$) | $\Delta E_{tot}$ ($E_0$) |
|---|---|---|---|---|---|
| terrace | 10 | $3s$ | 3 | -1 | |
| terrace | 11 | 0 | $3(s-1)$ | -2 | |
| terrace | 12 | $s(s-1)/2$ | $(s-1)(s-2)/2$ | -3 | |
| terrace | | | | | $-3(s+1)s/2$ |
| island | 5 | 3 | 3 | 7 | |
| island | 7 | $3(s-2)$ | $3(s-2)$ | 5 | |
| island | 9 | $(s-2)(s-3)/2$ | $(s-2)(s-3)/2$ | 3 | |
| island | | | | | $3(s+5)s/2$ |
| total | | | | | $6s$ |

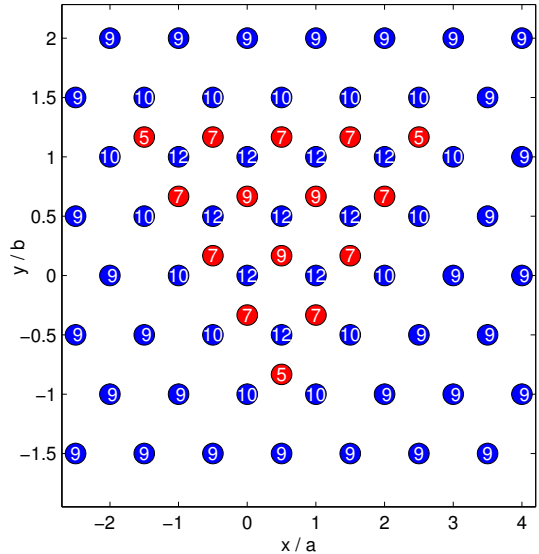

FIG. S.2.  Triangular island on c-plane, with A-step edges of length $s = 5$ atoms.  Terrace and island atoms are shown in blue and red, respectively, with $N_i$ value.

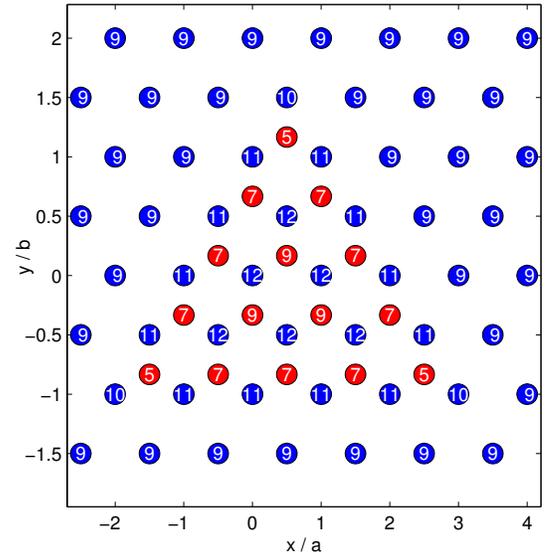

FIG. S.3.  Triangular island on c-plane, with B-step edges of length $s = 5$ atoms.  Terrace and island atoms are shown in blue and red, respectively, with $N_i$ value.

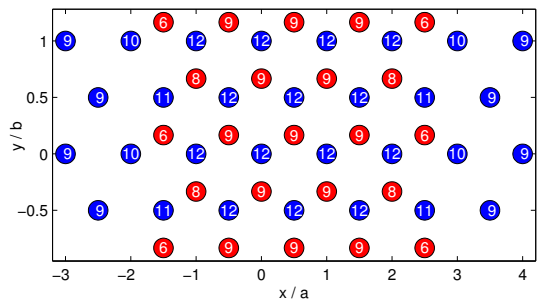

FIG. S.4. Top view of atomic structure of a linear island on the c-plane surface bounded by high-energy steps running normal to $\hat{x}$ direction. Terrace and island atoms are shown in blue and red, respectively, with $N_i$ value.

There are two structures for steps normal to $\hat{y}$ on the c-plane surface. They are similar to the "A" and "B" steps on a $(1\,1\,1)$ surface of a face-centered cubic crystal.[1] As we shall see below, they have the same step edge energy, but can have different ES step-edge barriers for diffusion of adatoms over the edge from above, depending upon the diffusion model.

A linear island with edges normal to $\hat{y}$ will be bounded by an A step on one side and a B step on the other. To evaluate the separate energies of the A and B steps, properly accounting for end effects, it is convenient to consider triangular islands with only A or B step boundaries, as shown in Figs. S.2 and S.3. Each atom is labeled with its number of nearest neighbors $N_i$. Atoms on the terrace layer near or under the island can have values greater than $N_i = 9$ for a normal surface atom, while atoms in the island can have values $N_i \leq 9$. The dependence of the total excess energy of the island on island size (characterized by its edge length of $s$ atoms) can be used to obtain the energy of its step edges. Table S.3 gives the formulas for the numbers of atoms $n_N$ in the island or nearby in the terrace, with various numbers of nearest neighbors $N_i$, as a function of $s$. These are converted into the island excess energy $\Delta E_{tot} = \sum_N n_N \Delta E_N$ through the excess energies $\Delta E_N = -(N_i - N_{ref})E_0$ of each terrace or island atom, relative to a reference value of $N_{ref} = 9$ or 12 for a terrace or island atom, respectively. The total energies are the same for islands with A- or B-step edges, indicating that the energies of the two types of steps are the same. The total excess island energy $6sE_0$ divided by the edge length $3s$ gives a step edge energy of $2E_0$ per unit length $a$ of step edge.

For steps normal to $\hat{x}$ on the c-plane surface, there is only one structure. A linear island with both edges having this structure is shown in Fig. S.4. Using a method similar to that above for extracting step edge energies from the $n_N$ values of atoms near the step, one obtains an edge energy of $4E_0$ per unit cell length $b$, which is $2.31E_0/a$ per unit length $a$. These steps thus have a higher energy than steps normal to $\hat{y}$.

To determine effective ES step edge barriers due to site geometry, we consider the sites that must be traversed by

a diffusing adatom, initially on the top of the island, to reach a low-energy site below the step edge.

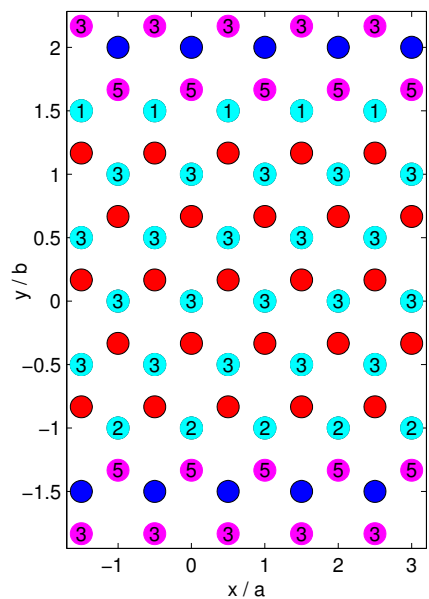

FIG. S.5. Top view of adatom structure of a linear island on the c-plane surface bounded by low-energy steps running normal to $\hat{y}$ direction. Upper edge at $y/b = 1$ is an A step, lower edge at $y/b = -1$ is a B step. Terrace and island atoms are shown in blue and red, while terrace and island adatom sites are shown in magenta and cyan, respectively, with $N_i$ value.

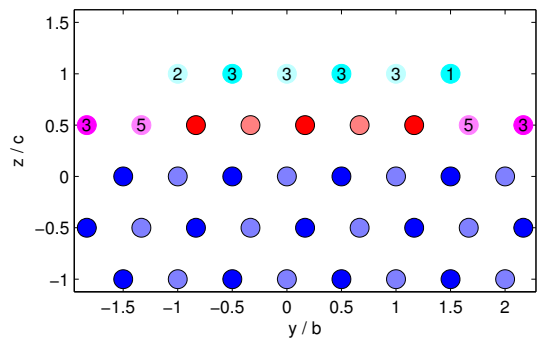

FIG. S.6. Side view of the adatom structure of low-energy steps on the c-plane surface shown in Fig. S.5. Atoms and adatom sites offset in $\hat{x}$, perpendicular to view plane, are shown in lighter colors.

Figures S.5 and S.6 show the adatom sites for steps normal to $\hat{y}$ with both A and B structures. The value of $N_i$ is indicated for each adatom site. To reach the low-energy $N_i = 5$ site below the step edge, an adatom on the island top with initial $N_i = 3$ diffusing between nearest-neighbor sites (NN kinetics) must traverse a higher-energy, lower-coordinated site. On the A step, the intermediate site has $N_i = 1$, while on the B step it has $N_i = 2$. Because we assign a very high energy to $N_i = 1$ atoms, the effective ES barrier for the A step approaches

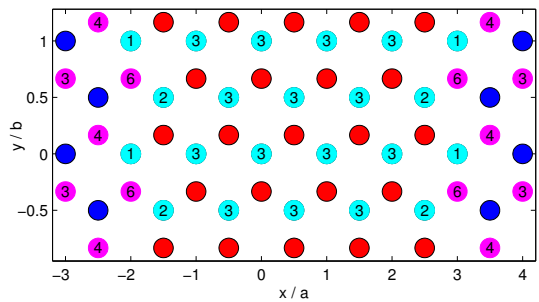

FIG. S.7. Top view of adatom structure of a linear island on the c-plane surface bounded by high-energy steps running normal to $\hat{x}$ direction. Terrace and island atoms are shown in blue and red, while terrace and island adatom sites are shown in magenta and cyan, respectively, with $N_i$ value.

infinity. For the B step, it is $\Delta E = 2E_0(N_j - N_i) = 2E_0$. In the NNN diffusion model, adatoms approaching the step edge can jump directly from the $N_i = 3$ site to the $N_i = 5$ site, bypassing the higher-energy site and giving an effective ES barrier of zero. For both A and B steps, NNN diffusion down the step edge occurs at the same rate as adatom diffusion on the island top.

Figure S.7 shows the adatom site structure for the steps normal to $\hat{x}$ direction. For NN kinetics, adatoms with $N_i = 3$ on the island top must traverse a site with $N_i = 2$ to reach the $N_i = 6$ site below the edge. This gives the same effective ES barrier obtained above for B steps. Likewise, for NNN kinetics, adatoms can jump directly from the $N_i = 3$ site to the $N_i = 6$ site, giving an effective ES barrier of zero.

### B. m-plane

For the m-plane surface, we considered step edges normal to $\hat{x}$, $\hat{z}$ and also $\hat{x} - \hat{z}$ (diagonal) directions.

There are two types of step edges normal to $\hat{z}$, which we label type A and type B. A linear island with both edges having the A structure is shown in Fig. S.8, while Fig. S.9 shows an island with B step edges. We evaluated the step edge energy as above, but in this case using two values of $N_{ref} = 8$ or 10 for alternating terrace atoms. For type A steps, the energy per unit cell length $a$ is $E_0$, while for type B steps it is $3E_0$.

There are also two types of steps normal to the $\hat{x}$ direction, which we label type A and type B. Linear islands with both edges having the same structure are shown in Figs. S.10 and S.12 (A step edges) and Fig. S.11 (B step edges). For type A steps, the energy per unit cell length $c$ is $3E_0$, while for type B steps it is $5E_0$. After accounting for the different unit cell lengths, these steps have higher energy per unit length than those normal to $\hat{z}$.

There is only a single structure for steps normal to the diagonal direction $\hat{x} - \hat{z}$. This structure, bounding the linear island shown in Fig. S.13, is also the same for steps normal to $\hat{x} + \hat{z}$. The energy per edge length $\sqrt{a^2 + c^2}$

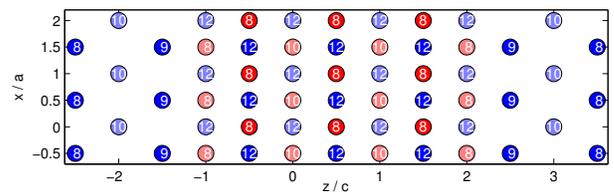

FIG. S.8. Top view of atomic structure of a linear island on the m-plane surface bounded by steps running normal to $\hat{z}$ direction, with step structure A. Terrace and island atoms are shown in blue and red, respectively, with $N_i$ value.

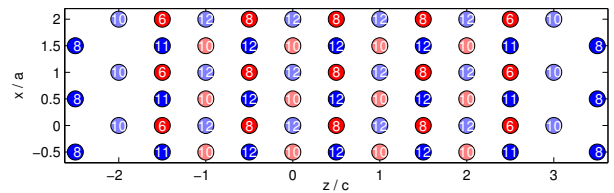

FIG. S.9. Top view of atomic structure of a linear island on the m-plane surface bounded by steps running normal to $\hat{z}$ direction, with step structure B. Terrace and island atoms are shown in blue and red, respectively, with $N_i$ value.

within a unit unit cell is $4E_0$, which gives a higher energy per unit length than A steps normal to either $\hat{z}$ or $\hat{x}$.

To analyze effective ES barriers at steps on the m-plane surface, one must remember that, for NN kinetics, there is a significant energy barrier to adatom diffusion in the $\hat{z}$ direction on the m-plane surface even with no steps present. This barrier of $\Delta E = 4E_0$ arises because of the rows of high-energy $N_i = 2$ adatom sites dividing the rows of low-energy $N_i = 4$ sites.

Figures S.14 and S.15 show the $N_i$ values for adatom sites on linear islands with A and B type step edges, respectively, normal to the $\hat{z}$ direction. For the A step, the adatom site at the top of the step has $N_i = 3$ rather than $N_i = 4$. In this case, the dominant barrier to diffusion down the step with NN kinetics remains the $N_i = 2$ adatom sites on the island top. So the effective ES barrier is $\Delta E = 4E_0$, the same as on a step-free surface. For the B step, the adatom site at the top of the step has $N_i = 1$ rather than $N_i = 4$. The ES barrier approaches infinity, because $N_i = 1$ atoms are assigned a very high energy. (Note that the $N_i = 6$ site below the step is at a lower level and is not a nearest neighbor of the $N_i = 2$ site above the step.) For NNN kinetics, the extra allowed jumps reduce the effective ES barriers to $\Delta E = 2E_0$ and $\Delta E = 4E_0$, respectively, for A and B steps. Both of these values are larger than the $\Delta E = 0$ for adatom diffusion on a step-free surface in the $\hat{z}$ direction with NNN kinetics.

Figures S.16 and S.17 show the adatom sites on linear islands with A and B type step edges, respectively, normal to the $\hat{x}$ direction. As shown in side view, Fig. S.18, step type A involves adatom sites with $N_i = 1$ at the step edge, producing an ES barrier approaching infinity for

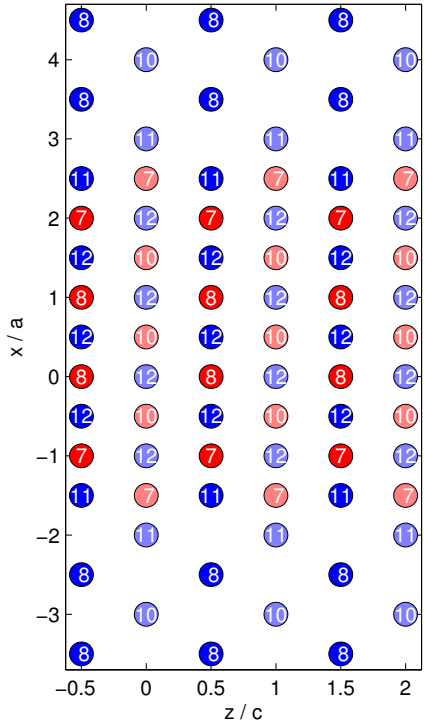

FIG. S.10. Top view of atomic structure of a linear island on the m-plane surface bounded by steps running normal to $\hat{x}$ direction, with step structure A. Terrace and island atoms are shown in blue and red, respectively, with $N_i$ value.

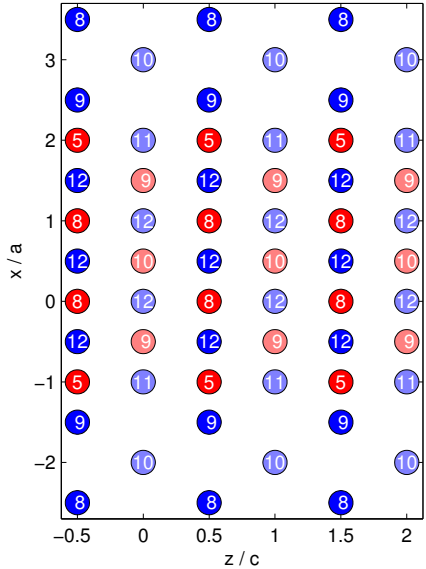

FIG. S.11. Top view of atomic structure of a linear island on the m-plane surface bounded by steps running normal to $\hat{x}$ direction, with step structure B. Terrace and island atoms are shown in blue and red, respectively, with $N_i$ value.

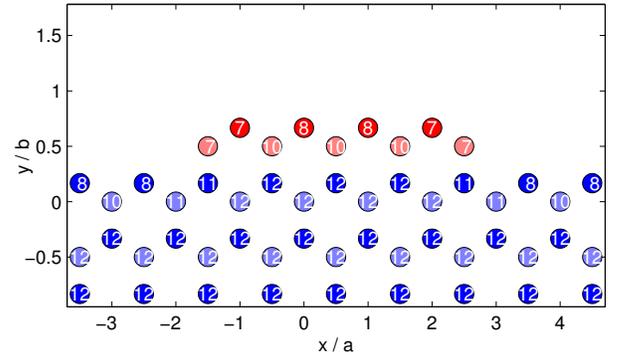

FIG. S.12. Side view of atomic structure of a linear island on the m-plane surface bounded by steps running normal to $\hat{x}$ direction, with step structure A. Terrace and island atoms are shown in blue and red, respectively, with $N_i$ value.

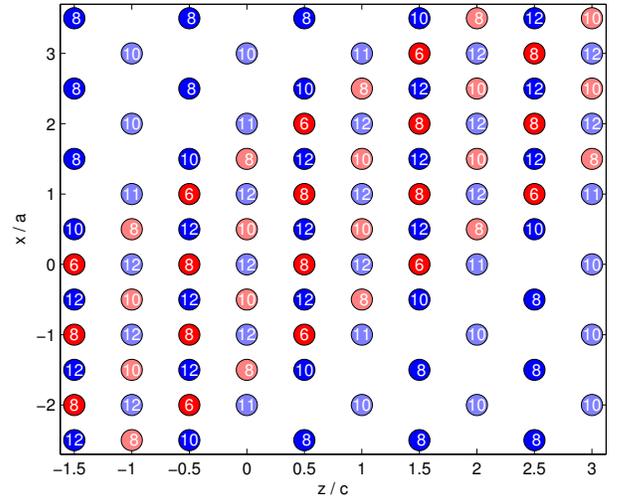

FIG. S.13. Top view of atomic structure of a linear island on the m-plane surface bounded by steps running normal to $\hat{x} - \hat{z}$ direction. Terrace and island atoms are shown in blue and red, respectively, with $N_i$ value.

NN kinetics. Diffusion down step type B only involves sites with $N_i = 3$, giving $\Delta E = 2E_0$ for NN kinetics. Both effective ES barriers are reduced to zero for NNN kinetics.

Figure S.19 shows the adatom sites on a linear islands with diagonal step edges, normal to the $\hat{x} - \hat{z}$ direction. The chains of low-energy $N_i = 4$ sites on the island top are blocked by an $N_i = 2$ site at the step edge, giving an ES barrier of $\Delta E = 4E_0$ for NN kinetics. This barrier is reduced to zero for NNN kinetics.

The step edge energies per unit length determined from this structural analysis have a significant orientation dependence. Those with lowest step edge energy are highlighted on Table S.2, and are discussed in the main paper. Use of NNN kinetics significantly reduces the effective ES barriers for almost all steps.

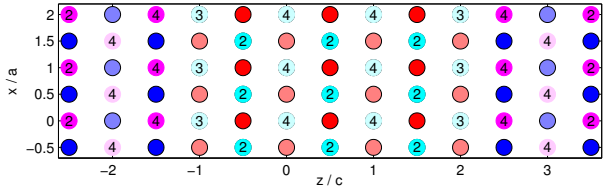

FIG. S.14. Top view of adatom structure of a linear island on the m-plane surface bounded by steps running normal to $\hat{z}$ direction, with step structure A. Terrace and island atoms are shown in blue and red, while terrace and island adatom sites are shown in magenta and cyan, respectively, with $N_i$ value.

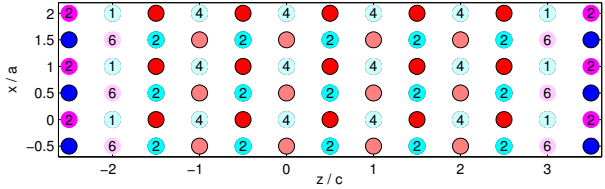

FIG. S.15. Top view of adatom structure of a linear island on the m-plane surface bounded by steps running normal to $\hat{z}$ direction, with step structure B. Terrace and island atoms are shown in blue and red, while terrace and island adatom sites are shown in magenta and cyan, respectively, with $N_i$ value.

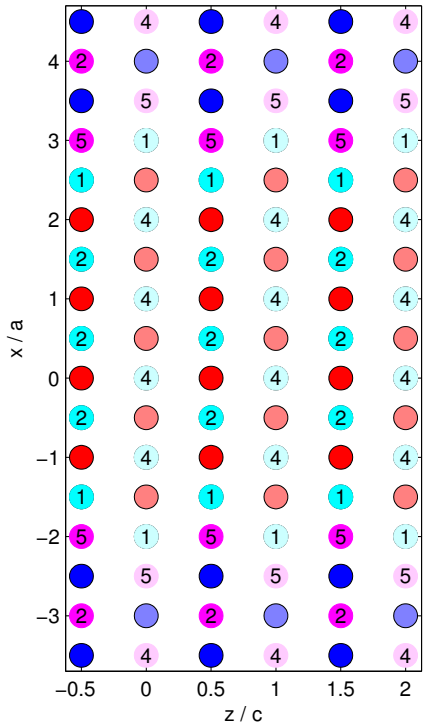

FIG. S.16. Top view of adatom structure of a linear island on the m-plane surface bounded by steps running normal to $\hat{x}$ direction, with step structure A. Terrace and island atoms are shown in blue and red, while terrace and island adatom sites are shown in magenta and cyan, respectively, with $N_i$ value.

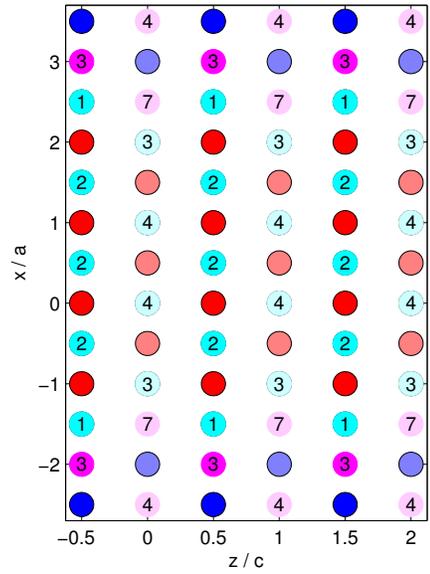

FIG. S.17. Top view of adatom structure of a linear island on the m-plane surface bounded by steps running normal to $\hat{x}$ direction, with step structure B. Terrace and island atoms are shown in blue and red, while terrace and island adatom sites are shown in magenta and cyan, respectively, with $N_i$ value.

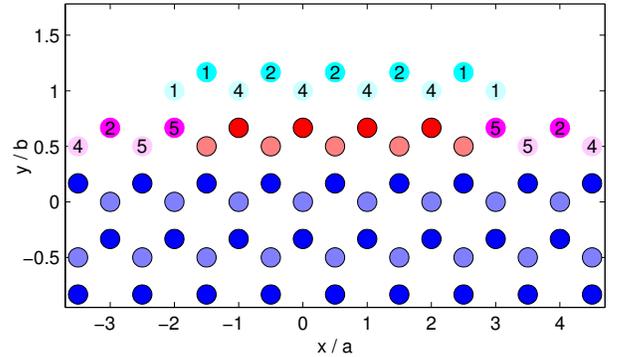

FIG. S.18. Side view of adatom structure of a linear island on the m-plane surface bounded by steps running normal to $\hat{x}$ direction, with step structure A. Terrace and island atoms are shown in blue and red, while terrace and island adatom sites are shown in magenta and cyan, respectively, with $N_i$ value.

## III. SURFACE DIFFUSION

Here we derive analytical formulas for diffusion coefficients of isolated adatoms in terms of the jump rates used in the KMC model, for different crystal surfaces. Simulations of single adatom diffusion using the KMC model were consistent with these formulas. The diffusion anisotropies calculated for the NN and NNN models can be compared with each other and with predictions from density functional theory (DFT) models of diffusion barriers.[2]

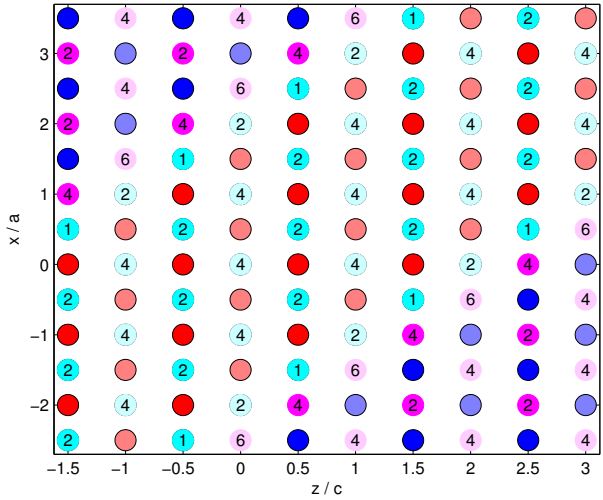

FIG. S.19. Top view of adatom structure of a linear island on the m-plane surface bounded by steps running normal to $\hat{x} - \hat{z}$ direction. Terrace and island atoms are shown in blue and red, while terrace and island adatom sites are shown in magenta and cyan, respectively, with $N_i$ value.

The diffusivity tensor $\mathbf{D}$ relates the diffusion flux $\vec{J}$ to the density gradient $\vec{\nabla}\rho$ through the expression

$$\vec{J} = -\mathbf{D} \cdot \vec{\nabla}\rho. \tag{S.1}$$

The component of the diffusion flux in a certain direction can be derived by considering the forward and reverse jump rates across a planar boundary normal to that direction. While derivations similar to this are given in many diffusion textbooks[3,4], what is new here are the cases for the m- and a-plane surfaces with sites having different energies and thus different jump rates. We first consider the simplest c-plane case to connect with textbook results.

At steady state in a system with a uniform density gradient, the flux across a planar boundary does not depend upon the position of the boundary, but just its orientation. If we break up a periodic lattice into surface unit cells, then the flux components normal to the unit cell boundaries can be obtained by considering a single unit cell. The flux component in the $x$ direction can be written as

$$J_x = \frac{1}{a_y} \sum_{\text{bdy}\perp x} (\theta_i \Gamma_{ij} - \theta_j \Gamma_{ji}), \tag{S.2}$$

where the sum is over all the nearest-neighbor jumps between sites $i$ and $j$ which cross the surface unit cell boundary perpendicular to $\hat{x}$, $\theta_i$ is the site occupancy fraction (the probability that an adatom site $i$ is occupied), $\Gamma_{ij}$ is the jump rate in the positive flux direction from site $i$ to site $j$, and $a_y$ is the length of the unit cell boundary normal to $\hat{x}$. (For simplicity, in this section we will use $\hat{x}$ and $\hat{y}$ as the two in-plane directions, as on the c-plane surface. Expressions for the m-and a-plane

surfaces can be obtained by appropriately permuting $\hat{x}$, $\hat{y}$ and $\hat{z}$.) A similar expression for the flux component in the $y$ direction can be obtained by swapping $x$ and $y$. The first term in the sum represents the forward jumps across the boundary, and the second term the reverse jumps. Here we assume non-interacting adatoms (small values of $\theta_i$, all neighboring sites $j$ and their neighbors unoccupied), so that the $\Gamma_{ij}$ are independent of the $\theta_i$.

To use this relation to obtain $\mathbf{D}$, we need to express the site occupancies in the neighborhood of the boundary in terms of the density gradient. At steady-state, the rate of jumps into each site is equal to the rate of jumps out of the site. For each site $i$, with neighboring sites $j$, one obtains

$$0 = \sum_j (\theta_i \Gamma_{ij} - \theta_j \Gamma_{ji}). \tag{S.3}$$

Here the sum is over all allowed jumps to or from site $i$, not just those crossing a surface unit cell boundary. Each difference $(\theta_i\Gamma_{ij} - \theta_j\Gamma_{ji})$ in the sum is proportional to the magnitude of the overall density gradient, and each is individually zero at equilibrium when there is no gradient or net flux. The jump rates and equilibrium site occupancies must satisfy detailed balance,

$$\theta_i^{eq} \Gamma_{ij} = \theta_j^{eq} \Gamma_{ji}, \tag{S.4}$$

where $\theta_i^{eq}$ is the equilibrium occupancy of site $i$ at zero flux. We can write the component of the density gradient in the $\alpha = x$ or $y$ direction as

$$\vec{\nabla}\rho|_\alpha = \frac{\Delta_\alpha}{a_\alpha}\rho, \tag{S.5}$$

where $\Delta_\alpha$ is a small parameter giving the fractional change in the density $\rho$ from unit cell to unit cell in the $\alpha$ direction. The density is obtained by summing the $\theta_i$ within a surface unit cell of area $a_x a_y$,

$$\rho = \frac{1}{a_x a_y} \sum_{\text{u.c.}} \theta_i. \tag{S.6}$$

In general, for small gradients, the site occupancies can be written as

$$\theta_i = \theta_i^{eq}(1 + g_{i\alpha}\Delta_\alpha), \tag{S.7}$$

where $g_{i\alpha}$ is the proportionality constant of site $i$ for gradients in the $\alpha$ direction. Using this relation and the detailed balance Eq.(S.4), the steady-state conditions Eq.(S.3) can be written

$$0 = \sum_j \Gamma_{ij}(g_{i\alpha} - g_{j\alpha}) \tag{S.8}$$

for each site $i$ and its neighbors $j$. It is only necessary for sites $i$ within a central unit cell to be considered. For neighboring sites $j$ outside of the central unit cell, the value of $g_{j\alpha}$ is incremented or decremented by unity if the

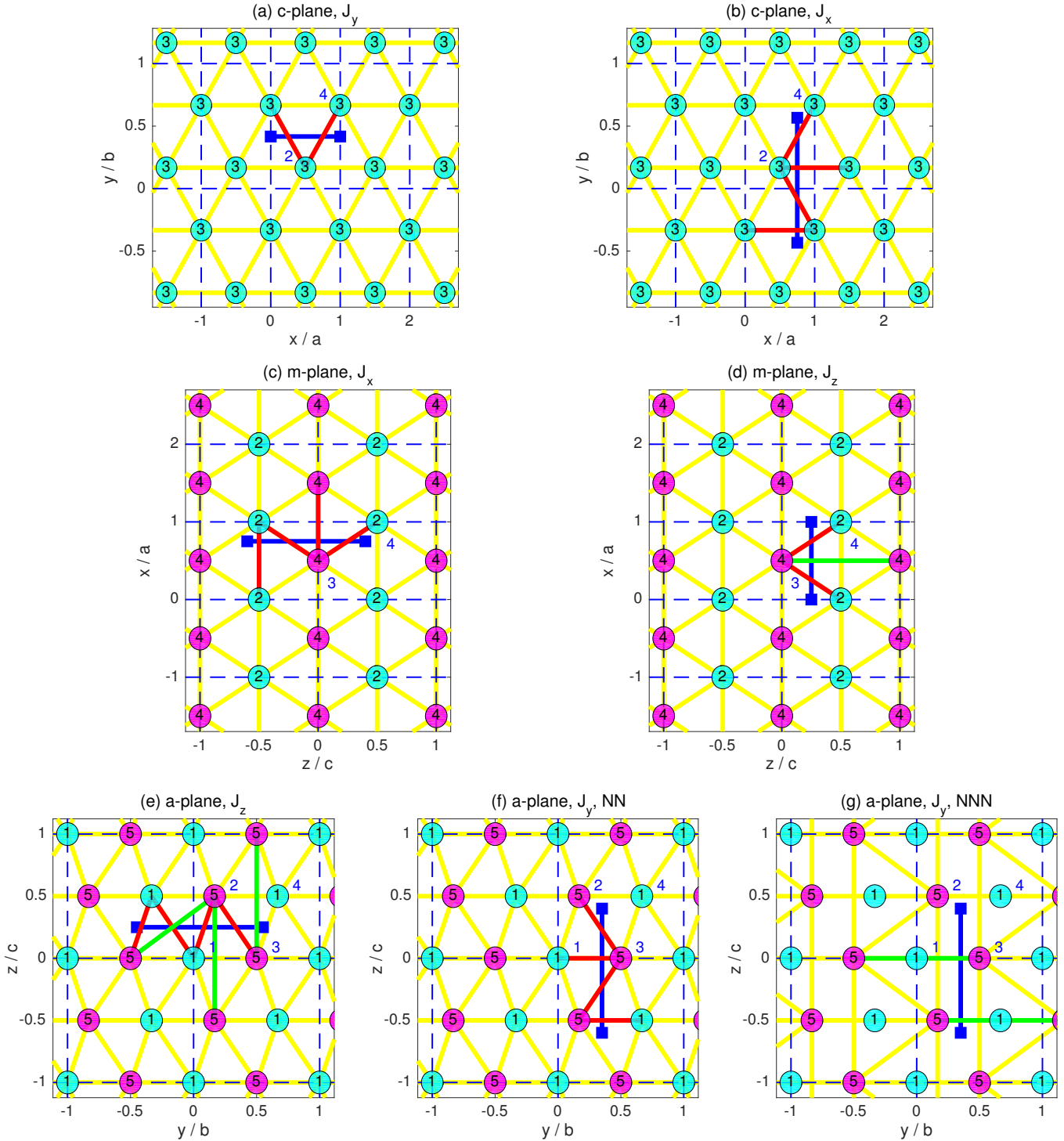

FIG. S.20. Schematics of adatom site arrangements and diffusion jumps on different crystal faces. (a) and (b) are for the c-plane surface, (c) and (d) are for the m-plane surface, and (e) - (g) are for the a-plane surface. Number $N_i$ of occupied nearest neighbors in surface below each site is given in circle marking site position. For m- and a-plane, magenta sites have higher $N_i$ (lower energy) than cyan sites. Dashed blue lines show surface unit cells. Yellow lines show nearest-neighbor jumps or, for (g), allowed next-nearest-neighbor jumps. Site numbers $i$ in a cell are shown in blue next to the site. Flux direction considered is vertical in (a), (c), and (e), and horizontal in (b), (d), (f), and (g). Red or green lines show NN or NNN jumps crossing a unit cell boundary segment perpendicular to the flux direction (bold blue line).

site is in a unit cell one step up or down the coordinate $\alpha$, respectively. This connection with the neighboring unit cells is what makes the $g_{i\alpha}$ values depend on the direction of the gradient, $\alpha$. Likewise, Eq.(S.2) for the flux $x$ component can be written as

$$J_x = \frac{1}{a_y} \sum_{\text{bdy} \perp x} \theta_{ij}^{eq} \Gamma_{ij}(g_{i\alpha} - g_{j\alpha})\Delta_\alpha, \qquad (S.9)$$

where $\alpha = x$ or $y$ gives the direction of the density gradient. Comparing this to the definition of $\mathbf{D}$, Eq.(S.1), and using Eqs.(S.5) and (S.6), the components of the diffusivity tensor can be written as

$$D_{\beta\alpha} = -\frac{a_\beta a_\alpha \sum_{\text{bdy} \perp \beta} \theta_i^{eq} \Gamma_{ij}(g_{i\alpha} - g_{j\alpha})}{\sum_{\text{u.c.}} \theta_i^{eq}}, \qquad (S.10)$$

where the components of the flux $\beta$ and the gradient $\alpha$ can be either the $x$ or $y$ directions. The problem now reduces to using the steady-state conditions, Eq.(S.8), to solve for the differences $g_{i\alpha} - g_{j\alpha}$ for various sites $i$, their neighbors $j$, and both gradient directions $\alpha = x$ or $y$, for use in the diffusivity expression Eq.(S.10). Below we calculate the diffusivity tensor components for the c- and m-plane structures.

Since the steady-state conditions, flux, and diffusivity expressions Eqs.(S.8,S.9,S.10) only involve differences in the $g$ coefficients, $g_{i\alpha} - g_{j\alpha}$, there is an arbitrary additive offset in the coefficients. This offset determines the position at which the density $\rho$ is equal to its equilibrium value $\sum_i \theta_i^{eq}$, even when a gradient is present. For convenience, we impose the condition

$$0 = \sum_{i \text{ in u.c.}} g_{i\alpha}, \qquad (S.11)$$

which makes the density $\rho$ equal to its equilibrium value in the central unit cell.

When the bond network on the surface is sufficiently symmetric (e.g. if there is a mirror plane perpendicular to the direction of the density gradient through every site), the site occupancies are simply related to the position $\vec{r_i}$ of the site along the density gradient. The coefficients $g_{i\alpha}$ can be written as

$$g_{i\alpha} = \frac{\vec{r_i} \cdot \hat{\alpha}}{a_\alpha}, \qquad (S.12)$$

and the diffusivity components as

$$D_{\beta\alpha} = \frac{a_\beta \sum_{\text{bdy} \perp \beta} \theta_i^{eq} \Gamma_{ij}(\vec{r_j} - \vec{r_i}) \cdot \hat{\alpha}}{\sum_{\text{u.c.}} \theta_i^{eq}}. \qquad (S.13)$$

The required symmetry is present on both the c-plane surface for gradients in either $\alpha = x$ or $y$ and m-plane for gradients in $\alpha = z$ or $x$. For the a-plane surface, mirror symmetry is present only perpendicular to $\hat{z}$. For gradients in $\alpha = y$, the a-plane surface must be treated with the more general expressions developed above. Note that the textbook formula[3] for $\mathbf{D}$ can be obtained from Eq. (S.13) when there is only one site per unit cell, and only one occupancy fraction and jump rate, which applies to the c-plane (but not the m- or a-plane).

## A. c-plane

The c-plane surface is simple because all sites have the same energy, and thus there is only one jump rate $\Gamma_0$ and equilibrium occupancy $\theta_0^{eq}$. Referring to Fig. S.20(a), the surface has lattice parameters $a_x = a$ and $a_y = b = \sqrt{3}a$, and all adatom sites have the same number of nearest neighbors in the surface below, $N_i = 3$. For diffusion in the $\beta = y$ direction, Fig. S.20(a), there are two jumps across the unit cell boundary segment (marked as bold red lines). The $\alpha = y$ component $(\vec{r_j} - \vec{r_i}) \cdot \hat{y}$ is $b/2$ in both cases. This gives

$$\begin{aligned}
D_{yy} &= \frac{b\theta_0^{eq}\Gamma_0(b/2 + b/2)}{2\theta_0^{eq}} \\
&= \frac{1}{2}\Gamma_0 b^2 \\
&= \frac{3}{2}\Gamma_0 a^2. \qquad (S.14)
\end{aligned}$$

The off-diagonal component $D_{yx}$ is zero because the $\alpha = x$ components $(\vec{r_j} - \vec{r_i}) \cdot \hat{x}$ of the two jumps are opposite and equal. For diffusion in the $\beta = x$ direction, as shown in Fig. S.20(b), there are four jumps across the unit cell boundary segment. The $\alpha = x$ component $(\vec{r_j} - \vec{r_i}) \cdot \hat{x}$ is $a/2$ for two and $a$ for the other two. This gives

$$\begin{aligned}
D_{xx} &= \frac{a\theta_0^{eq}\Gamma_0(a/2 + a/2 + a + a)}{2\theta_0^{eq}} \\
&= \frac{3}{2}\Gamma_0 a^2. \qquad (S.15)
\end{aligned}$$

The off-diagonal component $D_{xy}$ is zero because the $\alpha = y$ components $(\vec{r_j} - \vec{r_i}) \cdot \hat{y}$ of two of the jumps are opposite and equal, and the other two are zero. Since $D_{xx} = D_{yy}$, the diffusivity tensor is isotropic on the c-plane surface.

The results for isolated adatom diffusion on the c-plane surface do not vary between the NN and NNN diffusion models, because there are no next-nearest-neighbor sites of the adatom that fulfill the requirement regarding occupied and unoccupied intervening nearest neighbors in the NNN model.

## B. m-plane

### 1. NN kinetics

The m-plane surface is more complex because there are sites of two different energies, and so there are two jump rates and equilibrium occupancies. First we consider the NN diffusion model. Referring to Fig. S.20(c), an adatom on site 3 in any surface unit cell has $N_i = 4$, while an adatom on site 4 has $N_i = 2$ and thus a higher energy. For diffusion in the $\beta = x$ direction, Fig. S.20(c), there are four jumps across the unit cell boundary segment. Two of these jumps are "uphill" jumps and have a rate $\Gamma_u = f\Gamma_0$ that is lower than jumps on the c-plane by a factor of $f \equiv \exp(-4E_0/kT)$. The other two are

between sites of equal energy and have a rate $\Gamma_0$. The $\alpha = x$ component $(\vec{r_j} - \vec{r_i}) \cdot \hat{x}$ is $a/2$ for the first two and $a$ for the latter two. The equilibrium occupancy fraction of site 4 is lower than site 3 by a factor $f$. This gives

$$D_{xx} = \frac{a[\theta_0^{eq} f \Gamma_0 (a/2 + a/2) + \theta_0^{eq} \Gamma_0 a + f \theta_0^{eq} \Gamma_0 a)]}{\theta_0^{eq}(1+f)}$$

$$= \frac{(1+2f)}{(1+f)} \Gamma_0 a^2. \qquad (S.16)$$

where the factor $(1 + f)$ in the denominator comes from the sum of the different equilibrium occupancy fractions of the two sites in the unit cell. The off-diagonal component $D_{xz}$ is again zero because the $\alpha = z$ components $(\vec{r_j} - \vec{r_i}) \cdot \hat{z}$ of two of the jumps are opposite and equal, and the other two are zero. For diffusion in the $\beta = z$ direction, Fig. S.20(d), there are two jumps across the unit cell boundary segment. Both of these jumps are "uphill" jumps and have a rate $\Gamma_u$. The $\alpha = z$ component $(\vec{r_j} - \vec{r_i}) \cdot \hat{z}$ is $c/2$ in both cases, where $a_z = c = \sqrt{8/3}a$ is the unit cell length in the $z$ direction. This gives

$$D_{zz} = \frac{c\theta_0^{eq} f \Gamma_0 (c/2 + c/2)}{\theta_0^{eq}(1+f)}$$

$$= \frac{f}{(1+f)} \Gamma_0 c^2$$

$$= \frac{8f}{3(1+f)} \Gamma_0 a^2, \qquad (S.17)$$

The off-diagonal component $D_{zx}$ is zero because the $\alpha = x$ components $(\vec{r_j} - \vec{r_i}) \cdot \hat{x}$ of the two jumps are opposite and equal. Thus the model predicts anisotropic diffusion on the m-plane surface, with

$$\frac{D_{zz}}{D_{xx}} = \frac{8f}{3(1+2f)}. \qquad (S.18)$$

Because the factor $f$ depends upon temperature, the anisotropy varies strongly with temperature, as shown in Fig. S.21. For the m-plane with NN kinetics, the diffusivity is higher in the $x$ direction (perpendicular to [0 0 0 1]), because the chains of low energy sites align in this direction. This qualitatively agrees with the nearest-neighbor diffusion barrier anisotropy calculated using DFT.[2]

One can check the consistency of the m-plane and c-plane formulas because the connectivity of the sites is the same in both cases. If we use the m-plane results with the energy of the two sites made equal ($f = 1$), and the unit cell size in the $z$ direction set to $b = \sqrt{3}a$ rather than $c = \sqrt{8/3}a$, we obtain $D_{xx} = D_{zz} = (3/2)\Gamma_0 a^2$, which agrees with the c-plane results.

### 2. NNN kinetics

Next we consider the NNN diffusion model. The result for $D_{xx}$ will be the same for the NN and NNN models because the constraint of one unoccupied and one occupied intervening nearest neighbor is not satisfied. However,

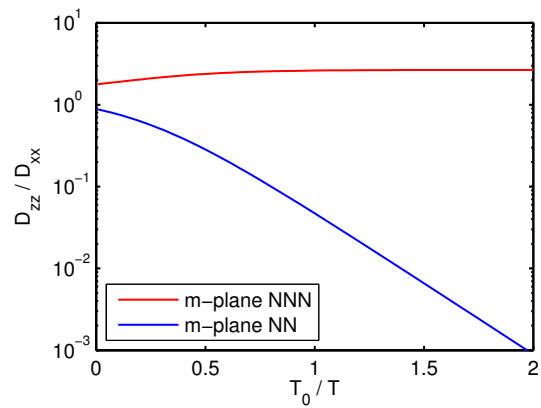

FIG. S.21. Comparison of predicted diffusion anisotropy $D_{zz}/D_{xx}$ vs. inverse temperature $T_0/T$ of the m-plane GaN surface, for the NN and NNN diffusion models.

for diffusion in the $z$ direction there is a possible next-nearest neighbor jump as shown in Fig. S.20(d), with the rate $\Gamma_0$ and jump length $c$. So for the NNN model on the m-plane surface, one obtains

$$D_{zz} = \frac{c[\theta_0^{eq} f \Gamma_0 (c/2 + c/2) + \theta_0^{eq} \Gamma_0 c]}{\theta_0^{eq}(1+f)}$$

$$= \Gamma_0 c^2$$

$$= \frac{8}{3} \Gamma_0 a^2, \qquad (S.19)$$

and the predicted diffusion anisotropy is

$$\frac{D_{xx}}{D_{zz}} = \frac{3(1+2f)}{8(1+f)}. \qquad (S.20)$$

The m-plane diffusion anisotropies for the NN and NNN models are shown as a function of temperature in Fig. S.21. For the NNN kinetics, diffusion is faster in the $z$ direction than the $x$ direction, which is opposite to the behavior of the NN kinetics.

### C. a-plane

For completeness, we derive the corresponding formulas for diffusion on the a-plane, even though we do not include other a-plane results in this paper. The a-plane surface is more complex than c- or m-plane, because there is no mirror symmetry perpendicular to the $\hat{y}$ direction. There are four sites per surface unit cell, and the lattice parameters are $a_y = b = \sqrt{3}a$ and $a_z = c = \sqrt{8/3}a$. Referring to Fig. S.20(e), sites 2 and 3 in each unit cell have $N_i = 5$, while sites 1 and 4 have $N_i = 1$ and thus a much higher energy. The equilibrium occupancy fraction of sites 1 and 4 is lower than that of sites 2 and 3 by a factor $f'$, and "uphill" jumps have a rate $\Gamma_u = f'\Gamma_0$ that is lower than that of the equal-energy or "downhill" jumps. If we use the non-linear dependence of $E_i$ on $N_i$, where atoms with $N_i = 1$ are assigned a very high energy,

then $f'$ approaches zero. For a linear $E_i(N_i)$, one obtains $f' = \exp(-8E_0/kT)$. Below we develop expressions first for NN kinetics and then NNN kinetics.

### 1. NN kinetics

For gradients in the $\alpha = z$ direction, the mirror symmetry allows us to use Eq.(S.13). For diffusion in the $\beta = z$ direction, Fig. S.20(e), there are four jumps across the unit cell boundary segment. The $\alpha = z$ component $(\vec{r_j} - \vec{r_i}) \cdot \hat{z}$ is $c/2$ in all cases. This gives

$$D_{zz} = \frac{c[\theta_0^{eq}\Gamma_0(1+f') + f'\theta_0^{eq}\Gamma_0(1+1)]c/2}{2\theta_0^{eq}(1+f')}$$
$$= \frac{(1+3f')}{4(1+f')}\Gamma_0 c^2$$
$$= \frac{2(1+3f')}{3(1+f')}\Gamma_0 a^2, \qquad (S.21)$$

where the factor $2(1+f')$ in the denominator comes from the sum of the different equilibrium occupancy fractions of the four sites in the unit cell. For diffusion in the $\beta = y$ direction, Fig. S.20(f), there are also four jumps across the unit cell boundary segment. Two have equal rates and opposite signs of their $\alpha = z$ components, while the other two have zero $z$ components, so the cross-term $D_{yz}$ is zero.

For gradients in the $\alpha = y$ direction, the mirror symmetry is not present, so we must use Eq.(S.10) rather than Eq.(S.13). The steady-state conditions Eq.(S.8) and the sum rule Eq.(S.11) must be used to solve for the $g_{iy}$ for the four sites $i = 1$ to 4. The steady-state conditions for the six nearest-neighbor jumps around each of the four sites give four equations,

$$6g_{1y} = (g_{3y} - 1) + 2(g_{4y} - 1) + 2g_{2y} + g_{3y},$$
$$(2+4f')g_{2y} = f'g_{4y} + 2f'g_{1y} + 2g_{3y} + f'(g_{4y} - 1),$$
$$(2+4f')g_{3y} = f'g_{1y} + 2g_{2y} + 2f'g_{4y} + f'(g_{1y} + 1),$$
$$6g_{4y} = (g_{2y} + 1) + 2g_{3y} + 2(g_{1y} + 1) + g_{2y}. \quad (S.22)$$

These can be rearranged to give

$$3 = -6g_{1y} + 2g_{2y} + 2g_{3y} + 2g_{4y},$$
$$f' = 2f'g_{1y} - (2+4f')g_{2y} + 2g_{3y} + 2f'g_{4y},$$
$$-f' = 2f'g_{1y} + 2g_{2y} - (2+4f')g_{3y} + 2f'g_{4y},$$
$$-3 = 2g_{1y} + 2g_{2y} + 2g_{3y} - 6g_{4y}. \quad (S.23)$$

These four equations are not independent, since the sum of the first and last is proportional to the sum of the middle two. By taking sums and differences of these equations, one obtains the three independent relations

$$g_{3y} - g_{2y} = \frac{f'}{2(1+f')},$$
$$g_{4y} - g_{1y} = 3/4,$$
$$g_{3y} + g_{2y} = g_{4y} + g_{1y}. \quad (S.24)$$

Using the sum rule Eq.(S.11), one obtains

$$g_{1y} = -3/8,$$
$$g_{2y} = \frac{-f'}{4(1+f')},$$
$$g_{3y} = \frac{f'}{4(1+f')},$$
$$g_{4y} = 3/8. \qquad (S.25)$$

For diffusion in the $\beta = y$ direction, substituting these into Eq.(S.10) and using the NN boundary jumps shown in Fig. S.20(f) gives

$$\frac{\sum_{\text{bdy}\perp\beta}\theta_i^{eq}\Gamma_{ij}(g_{i\alpha} - g_{j\alpha})}{\theta_0^{eq}\Gamma_0}$$
$$= f'(g_{2y} - g_{4y}) + 2(g_{2y} - g_{3y}) + f'(g_{1y} - g_{3y})$$
$$= f'(g_{1y} - g_{4y}) + (2+f')(g_{2y} - g_{3y})$$
$$= -f'\left(\frac{3}{4} + \frac{2+f'}{2(1+f')}\right), \qquad (S.26)$$

$$D_{yy} = \frac{b^2\theta_0^{eq}\Gamma_0 f'\left(\frac{3}{4} + \frac{2+f'}{2(1+f')}\right)}{2\theta_0^{eq}(1+f')}$$
$$= \frac{f'(7+5f')}{8(1+f')^2}\Gamma_0 b^2$$
$$= \frac{3f'(7+5f')}{8(1+f')^2}\Gamma_0 a^2. \qquad (S.27)$$

For diffusion in the $\beta = z$ direction, using the NN boundary jumps shown in Fig. S.20(e) gives

$$\frac{\sum_{\text{bdy}\perp\beta}\theta_i^{eq}\Gamma_{ij}(g_{i\alpha} - g_{j\alpha})}{\theta_0^{eq}\Gamma_0}$$
$$= f'(g_{1y} - g_{2y}) + (g_{3y} - g_{2y}) + f'(g_{3y} - g_{4y})$$
$$+ f'(g_{1y} + 1 - g_{4y})$$
$$= 2f'(g_{1y} - g_{4y}) + (1+f')(g_{3y} - g_{2y}) + f'$$
$$= -3f'/2 + f'/2 + f'$$
$$= 0, \qquad (S.28)$$

so the off-diagonal component $D_{zy}$ is again zero. Here as in Eqs.(S.22) we have incremented or decremented the values of $g_{iy}$ given in Eqs.(S.25) by unity when the site is one unit cell up or down the density gradient. The model predicts anisotropic diffusion on the a-plane surface, with

$$\frac{D_{yy}}{D_{zz}} = \frac{9f'(7+5f')}{16(1+3f')(1+f')}. \qquad (S.29)$$

Because the factor $f'$ depends upon temperature, the anisotropy varies strongly with temperature, as shown in Fig. S.22. For the a-plane, the diffusivity is higher in the $z$ direction (along [0001]), because the chains of low energy sites align in this direction.

One can check the consistency of the a-plane and c-plane formulas because the connectivity of the sites is the same in both cases. For the $\hat{z}$ direction, the a-plane

result with $f' = 1$ is $D_{zz} = (1/2)\Gamma_0 c^2$. If we adjust the unit cell size $c$ to $b$, this agrees with the c-plane result. For the $\hat{y}$ direction, the a-plane result with $f' = 1$ is $D_{yy} = (3/8)\Gamma_0 b^2$. If we adjust the unit cell size $b$ to $2a$ (since the a-plane has twice the number of sites per unit cell as the c-plane), this agrees with the c-plane result.

### 2. NNN kinetics

For NNN kinetics, additional jumps are allowed. For gradients in the $\alpha = z$ direction we can use Eq.(S.13). For diffusion in the $\beta = z$ direction, Fig. S.20(e) shows the two additional jumps of length $c$ and one with a $z$ component of $c/2$. Adding these gives

$$D_{zz} = \frac{c[\theta_0^{eq}\Gamma_0(6 + f') + f'\theta_0^{eq}\Gamma_0(1 + 1)]c/2}{2\theta_0^{eq}(1 + f')}$$
$$= \frac{3(2 + f')}{4(1 + f')}\Gamma_0 c^2$$
$$= \frac{2(2 + f')}{(1 + f')}\Gamma_0 a^2. \tag{S.30}$$

For diffusion in the $\beta = y$ direction, Fig. S.20(g) shows that the two additional jumps have zero component in the $y$ direction, so $D_{yz} = 0$ as for NN kinetics.

For gradients in the $\alpha = y$ direction, we have to recalculate the $g_{iy}$ including the additional allowed next-nearest-neighbor jumps between sites 2 and 3 shown in yellow in Fig. S.20(g). The four steady-state conditions become

$$6g_{1y} = (g_{3y} - 1) + 2(g_{4y} - 1) + 2g_{2y} + g_{3y},$$
$$(4 + 4f')g_{2y} = f'(2g_{4y} - 1) + 2f'g_{1y} + 2(2g_{3y} - 1),$$
$$(4 + 4f')g_{3y} = f'(2g_{1y} + 1) + 2(2g_{2y} + 1) + 2f'g_{4y},$$
$$6g_{4y} = (g_{2y} + 1) + 2g_{3y} + 2(g_{1y} + 1) + g_{2y}. \tag{S.31}$$

These can be rearranged to give

$$3 = -6g_{1y} + 2g_{2y} + 2g_{3y} + 2g_{4y},$$
$$f' + 2 = 2f'g_{1y} - (4 + 4f')g_{2y} + 4g_{3y} + 2f'g_{4y},$$
$$-(f' + 2) = 2f'g_{1y} + 4g_{2y} - (4 + 4f')g_{3y} + 2f'g_{4y},$$
$$-3 = 2g_{1y} + 2g_{2y} + 2g_{3y} - 6g_{4y}. \tag{S.32}$$

Taking sums and differences of these equations gives the three independent relations

$$g_{3y} - g_{2y} = 1/2,$$
$$g_{4y} - g_{1y} = 3/4,$$
$$g_{3y} + g_{2y} = g_{4y} + g_{1y}. \tag{S.33}$$

Using the sum rule Eq.(S.11), one obtains

$$g_{1y} = -3/8,$$
$$g_{2y} = -1/4,$$
$$g_{3y} = 1/4,$$
$$g_{4y} = 3/8. \tag{S.34}$$

For diffusion in the $\beta = y$ direction, substituting these into Eq.(S.10) and using the NN and NNN boundary jumps shown in Figs. S.20(f) and S.20(g) gives

$$\frac{\sum_{\text{bdy}\perp\beta}\theta_i^{eq}\Gamma_i(g_{i\alpha} - g_{j\alpha})}{\theta_0^{eq}\Gamma_0}$$
$$= f'(g_{2y} - g_{4y}) + 2(g_{2y} - g_{3y}) + f'(g_{1y} - g_{3y}) - 2$$
$$= f'(g_{1y} - g_{4y}) + (2 + f')(g_{2y} - g_{3y}) - 2$$
$$= -(12 + 5f')/4, \tag{S.35}$$

$$D_{yy} = \frac{b^2\theta_0^{eq}\Gamma_0(12 + 5f')/4}{2\theta_0^{eq}(1 + f')}$$
$$= \frac{12 + 5f'}{8(1 + f')}\Gamma_0 b^2$$
$$= \frac{3(12 + 5f')}{8(1 + f')}\Gamma_0 a^2, \tag{S.36}$$

For diffusion in the $\beta = z$ direction, using the NN and NNN boundary jumps shown in Fig. S.20(e) gives

$$\frac{\sum_{\text{bdy}\perp\beta}\theta_i^{eq}\Gamma_{ij}(g_{i\alpha} - g_{j\alpha})}{\theta_0^{eq}\Gamma_0}$$
$$= f'(g_{1y} - g_{2y}) + (g_{3y} - g_{2y}) + f'(g_{3y} - g_{4y})$$
$$+ f'(g_{1y} + 1 - g_{4y}) + [g_{3y} - (g_{2y} + 1)]$$
$$= 2f'(g_{1y} - g_{4y}) + (2 + f')(g_{3y} - g_{2y}) + f' - 1$$
$$= -3f'/2 + (2 + f')/2 + f' - 1$$
$$= 0, \tag{S.37}$$

so $D_{zy} = 0$ as for NN kinetics. The model predicts anisotropic diffusion on the a-plane surface, with

$$\frac{D_{yy}}{D_{zz}} = \frac{3(12 + 5f')}{16(2 + f')}. \tag{S.38}$$

Figure S.22 shows the adatom diffusion anisotropy on the a-plane as a function of inverse temperature, for both the NN and NNN diffusion models, using the linear formula for $E_i(N_i)$. For the a-plane, the diffusivity is higher in the $\hat{z}$ direction with NN kinetics, because the chains of low-energy sites align in this direction. This anisotropy qualitatively agrees with the nearest-neighbor diffusion barrier anisotropy calculated using DFT.[2] With NNN kinetics, surface diffusion is almost isotropic. For the non-linear $E_i(N_i)$ formula, with $f' = 0$, the diffusion anisotropy approaches the low-temperature limit of the linear formula, with $D_{yy}/D_{zz} = 0$ for NN kinetics and $D_{yy}/D_{zz} = 9/8$ for NNN kinetics.

## IV. IMPLEMENTATION IN SPPARKS

We carried out the KMC simulations using the Stochastic Parallel PARticle Kinetic Simulator (SP-PARKS) computer code.[5,6] The dynamics were calculated using the variable time step method known as the

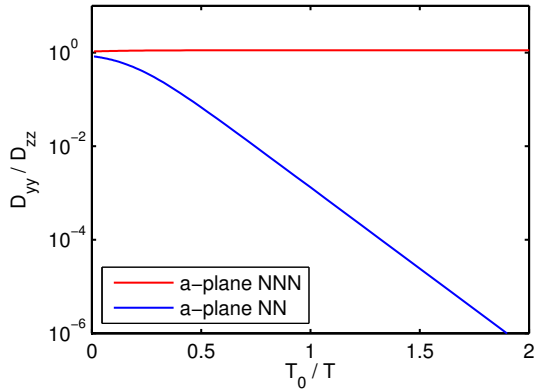

FIG. S.22. Comparison of predicted diffusion anisotropy $D_{yy}/D_{zz}$ vs. inverse temperature $T_0/T$ of the a-plane GaN surface, for the NN and NNN diffusion models, using the linear $E_i(N_i)$ relationship.

Gillespie or BKL algorithm.[7,8] The SPPARKS code implements 3-dimensional lattice site simulations using orthorhombic unit cells.

For the c-plane surface, the outward surface normal is in the $z$ direction, and periodic boundary conditions are applied in the $x$ and $y$ directions. The simulation cell size was $128a$ by $74b$ to give similar overall dimensions in the $x$ and $y$ directions, respectively. For m-plane, $y$ is the surface normal and the simulation cell is periodic in $z$ and $x$. The simulation cell size was $78c$ by $128a$ in the $z$ and $x$ directions, respectively. In the surface normal direction, the simulation box length was typically 12 unit cells. The lowest five unit cells were initially fully occupied, with the sites above unoccupied, to represent the interface between the crystal substrate and the environment. These sizes were chosen so that the crystal surface did not reach either boundary during the time scale of the simulations.

A deposition event is simulated by choosing a random in-plane position, and placing an atom into the closest eligible unoccupied site. Eligible unoccupied sites must have $N_i$ in the range $2 \le N_i \le 11$. If more than one has the same in-plane distance from the chosen position, the atom is placed into the site with highest height. The deposition rate is an input parameter that defines the average number of deposition events per unit simulation time. The growth rate $G$ in monolayers per unit simulation time (ML/ut) is calculated by dividing the deposition rate, which is an input parameter, by the number of sites per monolayer in the simulation box. For the results presented here, the number of sites per monolayer was $2 \times 128 \times 74 = 18944$ or $2 \times 78 \times 128 = 19968$ for c- or m-plane, respectively.

We investigated potential effects of the finite simulation size. We expect finite-size effects to arise when the adatom diffusion lengths or island spacings become comparable to the system size, that is, at low growth rates and high temperatures. We studied finite-size effects for

TABLE S.4. The effect of simulation size on island spacing $S/a$ and oscillation amplitude $\Delta$, for a deposition rate of $G = 1.3 \times 10^{-6}$ ML/ut and temperature range $T/T_0$ from 0.2 to 0.6 for m-plane with NNN diffusion kinetics.

| size | $T/T_0$ | 0.2 | 0.3 | 0.4 | 0.5 | 0.6 |
|---|---|---|---|---|---|---|
| $78c \times 128a$ | $S/a$ | 10.4 | 15.6 | 23.5 | 36.8 | 56.7 |
| $156c \times 256a$ | $S/a$ | 11.4 | 17.3 | 28.2 | 42.5 | 62.8 |
| $78c \times 128a$ | $\Delta$ | 0.21 | 0.28 | 0.47 | 0.74 | 0.87 |
| $156c \times 256a$ | $\Delta$ | 0.21 | 0.28 | 0.55 | 0.85 | 0.92 |

TABLE S.5. Typical number of KMC events and computational time for simulation of growth of 5.1 ML on the m-plane at a deposition rate of 0.1 de/ut with NN or NNN diffusion kinetics and different temperatures.

| Diffusion Model | $T/T_0$ | # Events | Comp. Time (s) | Time / Event (s) |
|---|---|---|---|---|
| NN | 0.3 | $1.08 \times 10^8$ | $1.91 \times 10^4$ | $1.77 \times 10^{-4}$ |
| NN | 0.4 | $8.89 \times 10^7$ | $1.36 \times 10^4$ | $1.53 \times 10^{-4}$ |
| NN | 0.5 | $1.05 \times 10^8$ | $1.56 \times 10^4$ | $1.48 \times 10^{-4}$ |
| NN | 0.6 | $1.46 \times 10^8$ | $2.11 \times 10^4$ | $1.44 \times 10^{-4}$ |
| NNN | 0.3 | $3.84 \times 10^7$ | $2.37 \times 10^4$ | $6.17 \times 10^{-4}$ |
| NNN | 0.4 | $3.71 \times 10^7$ | $2.17 \times 10^4$ | $5.85 \times 10^{-4}$ |
| NNN | 0.5 | $6.30 \times 10^7$ | $3.49 \times 10^4$ | $5.54 \times 10^{-4}$ |
| NNN | 0.6 | $1.22 \times 10^8$ | $6.39 \times 10^4$ | $5.24 \times 10^{-4}$ |

a low growth rate of $G = 1.3 \times 10^{-6}$ ML/ut and temperatures $T/T_0$ from 0.2 to 0.6 for the m-plane with NNN kinetics. Table S.4 shows the effect of doubling the system size on the values of island spacing $S/a$ and oscillation amplitude $\Delta$. Both values tend to increase when the system size increases. This indicates that the $S/a$ and $\Delta$ values given in the main paper at low $G$ and high $T$, using the fixed smaller simulation size, are likely depressed by $\sim10\%$ due to finite-size effects. However, we expect the trend of the growth mode transitions will not significantly change.

Table S.5 gives typical values for the total number of KMC events (diffusion jumps or deposition events) and the required computational time for some specific simulations using Argonne Laboratory Computing Resource Center (LCRC) computing resources. As temperature increases, the event rate is generally increasing because there is more diffusion at high temperature. There are fewer events in NNN than NN at the same temperature, while the computation time is larger in the NNN kinetics. At low temperature, the diffusion jump rate is lower and the fraction of deposition events is relatively high. Since deposition events require more computation time, the time per event is larger at lower temperature.

## V. SUPPLEMENTAL SIMULATION RESULTS

### A. Equilibrium occupation

Here we provide additional discussion of the equilibrium occupation fractions of different sites on the m-plane surface, and how they affect the occupations at high temperature during LBL growth, Fig. 3(d) in the main paper. We have found that, for both c-plane and m-plane, the equilibrium occupation fractions $\theta_i^{eq}$ in the adatom and surface layers follow closely the relation similar to Langmuir isotherm

$$\frac{\theta_i^{eq}}{1-\theta_i^{eq}} = \exp\left[(2N_i-12)T_0/T\right].\qquad\text{(S.39)}$$

Here the Boltzmann factor represents the energy $-2N_iE_0$ of the occupied site, relative to the average energy $-12E_0$. For the m-plane, adatoms in the lower or higher sites of each monolayer have $N_i = 4$ or 2. Thus at $T = T_0$, the equilibrium occupancy of adatoms in the $N_i = 4$ sites (1 or 3 in Fig. 1 of the main paper) becomes appreciable, $\theta_i^{eq} = 0.018$. At the beginning of growth in Fig. 3(d), site 3 (blue curve) already has this occupancy. As the first monolayer grows, and the occupancy of sites 3 and 4 (blue and green curves) ramp up linearly toward unity, the occupancy of adatoms in site 1 in the next monolayer (red curve) ramps up to its equilibrium value. This is also reflected in the vacancy concentrations in the nearly-filled layers. The lower or higher sites at the surface of a filled monolayer have $N_i = 10$ or 8. The equilibrium vacancy concentration in the $N_i = 8$ sites 2 or 4 can be appreciable, $1-\theta_i^{eq} = 0.018$ at $T = T_0$, mirroring the adatom occupation. As each monolayer fills in turn, these sites saturate at an occupation near 0.98, and the remaining vacancies become occupied during the growth of the subsequent monolayer.

### B. Reciprocal Space Intensity

Here we calculate the intensity in reciprocal space measured in x-ray scattering experiments, in terms of reciprocal space coordinates $HKL$ of the orthohexagonal unit cells.[9] These can be related to the Miller-Bravais indices $hkil$ of the hexagonal GaN unit cell by $h = H$, $k = (-H + K)/2$, $i = (-H - K)/2$, and $l = L$.

The scattered intensity is proportional to the square of the complex structure factor $F$, which we calculate as the sum of terms from the simulation volume and from a semi-infinite crystal substrate located beneath the simulation volume.

$$I(H,K,L) \propto |F_{sim}+F_{sub}|^2 \qquad\text{(S.40)}$$

The contribution from the simulation volume $F_{sim}$ is simply the sum of the scattering factor $f$ from each atom, multiplied by a phase factor $\exp(i\mathbf{Q}\cdot\mathbf{r}_j)$, where $\mathbf{Q} \equiv 2\pi(H/a, K/b, L/c)$ is the scattering wavevector and

$\mathbf{r}_j$ is the position of atom $j$. Since the simulation has periodic boundary conditions, this can be conveniently written as a discrete Fourier transform in the in-plane directions,

$$F_{sim} = \sum_{j\ occupied} f\exp(i\mathbf{Q}\cdot\mathbf{r}_j) = FFT\left(A(r_{j1},r_{j2})\right),\qquad\text{(S.41)}$$

where the complex amplitude $A$ is obtained by summing over the column of atoms at each in-plane location $(r_{j1}, r_{j2})$,

$$A(r_{j1},r_{j2}) = \sum_{j\ in\ column} f\exp(i\mathbf{Q}\cdot r_{j3}),\qquad\text{(S.42)}$$

and $r_{j1}$ and $r_{j2}$ are the in-plane components and $r_{j3}$ is the out-of-plane component of $\mathbf{r}_j$. The out-of-plane component is in the $z$ direction for the c-plane surface, or the $y$ direction for the m-plane surface.

The substrate only contributes along the crystal truncation rods (CTRs)[10] extending normal to the surface through the Bragg peaks at integer values of the in-plane reciprocal space coordinates ($H$ and $K$ for c-plane, $L$ and $H$ for m-plane), and has the form

$$F_{sub} = \frac{f_{uc}n_1 n_2 \exp(-iQ_3 a_3)}{1-\exp(-iQ_3 a_3)}.\qquad\text{(S.43)}$$

Here $a_3$ is the lattice parameter normal to the surface, $n_1$ and $n_2$ are the number of units cells within the periodic boundary conditions of the simulation, and $f_{uc}$ is the structure factor of a unit cell,

$$f_{uc} = \sum_{uc} f\exp[2\pi i(Hx_j + Ky_j + Lz_j)],\qquad\text{(S.44)}$$

where $x_j$, $y_j$, and $z_j$ are the fractional coordinates of the site $j$ in a unit cell. Because of cancellations between the multiple sites in the orthohexagonal unit cell, for the c-plane surface $H$ and $K$ must be either both even or both odd for $f_{uc}$ to be non-zero, corresponding to a CTR.

### C. m-plane with NN Kinetics

Here we present some additional results, for m-plane growth using NN kinetics, for comparison with parallel results in the main paper on c-plane growth using NN kinetics and m-plane growth using NNN kinetics. We extract the boundaries between the 3-dimensional (3D), layer-by-layer (LBL), and step-flow (SF) homoepitaxial growth modes.

#### 1. LBL-to-3D transition

Figure S.23 shows the behavior of the normalized anti-Bragg intensity oscillation amplitude $\Delta$ as function of the growth rate and inverse temperature for m-plane with

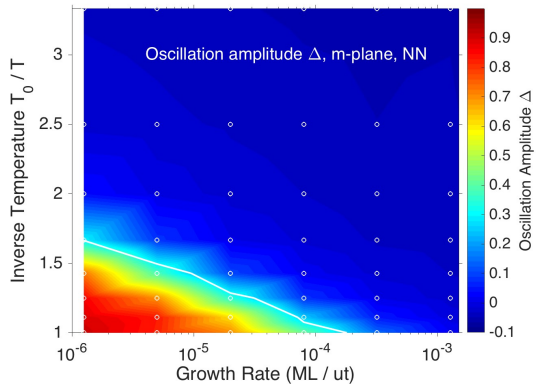

FIG. S.23. Plots of $\Delta$ as a function of inverse temperature and growth rate for m-plane with NN kinetics. The white circles show the conditions for the simulation data points, and the color scale gives the interpolated value of $\Delta$. The white contour at $\Delta = 0.3$ indicates the LBL-3D boundary.

NN kinetics. The white contour line is the LBL-to-3D boundary at $\Delta = 0.3$. We fit this boundary to the expression given in the main paper, and obtained the parameters $E_{3D}/E_0 = 6.83 \pm 0.04$ and $\log_{10} A_{3D}(\text{ML/ut}) = -0.88 \pm 0.02$, for m-plane with NN kinetics. Compared to m-plane with NNN kinetics, the LBL range is smaller. However, it is still larger than c-plane with NN kinetics.

### 2. Island spacing

The average island spacing is given by $S = 2\pi/\overline{Q}$ at 0.5 ML, where $\overline{Q}$ is here extracted from the diffuse scattering distribution in the $H$ direction, as shown in Fig. S.24. The values of $S$ obtained are plotted as function of inverse temperature and growth rate for m-plane with NN kinetics in Fig. S.25. We have fit the island spacings for higher temperatures and lower growth rates to the expression given in the main paper. Figure S.26 shows the results of the fit. The fitting parameters obtained are $n = 0.266 \pm 0.004$, $E_S/E_0 = 1.24 \pm 0.06$ and $\log_{10} G_S(\text{ML/ut}) = 1.92 \pm 0.06$. Compared to the results for m-plane with NNN kinetics and c-plane with NN kinetics, the power-law slope is similar, $n \approx -1/4$, while the apparent activation energy $E_S/E_0$ is significantly different.

### 3. Growth mode map

From the fits to the island spacing, the SF-to-LBL boundary can be estimated, as described in the main paper. The predicted growth mode map including both the LBL-to-3D and SF-to-LBL growth mode transitions is shown in Fig. S.27. The red squares are the simulation data for the LBL-to-3D boundary. Overall, the map resembles that for c-plane with NN kinetics more than

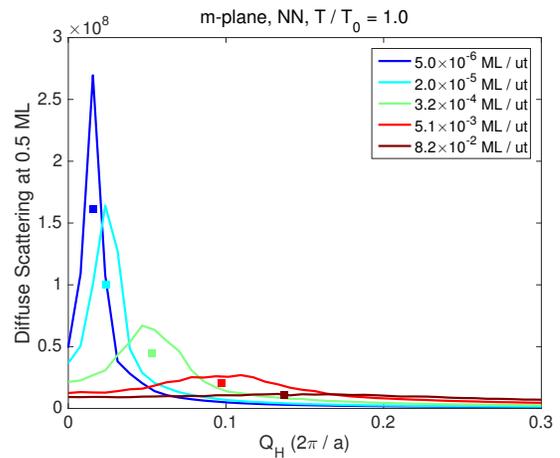

FIG. S.24. Diffuse scattering profiles near the CTR at 0.5 ML for different growth rates given in legend. Results are shown for m-plane, $T/T_0 = 1.0$, NN kinetics. Square symbols show the extracted position $\overline{Q}$.

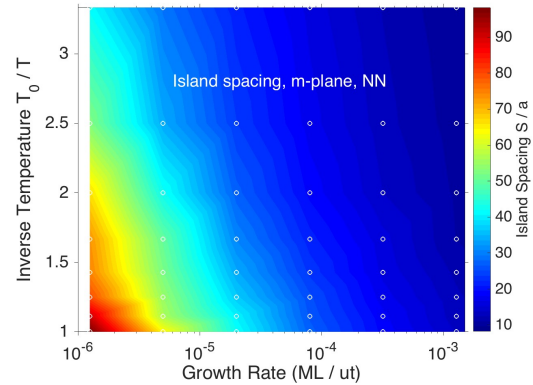

FIG. S.25. Island spacing $S/a$ at 0.5 ML as a function of inverse temperature and growth rate for m-plane with NN kinetics. The white circles show the conditions for the simulation data points, and the color scale gives the interpolated value of $S/a$.

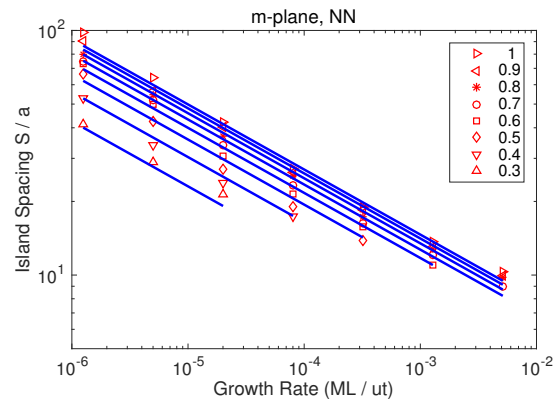

FIG. S.26. Island spacing $S/a$ as a function of growth rate at various $T/T_0$ given in the legend for m-plane with NN kinetics. Lines are fits of all points to expression given in main paper.

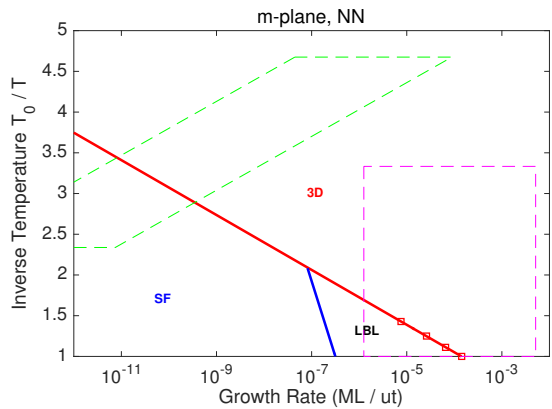

FIG. S.27. Predicted homoepitaxial growth mode map for GaN m-plane with NN kinetics. Red and blue lines show the boundaries between 3-dimensional (3D), layer-by-layer (LBL), and step-flow (SF) growth modes, for a terrace width of $W/a = 125$. Magenta and green dashed boundaries show ranges of conditions for simulations and experiments, respectively.

that for m-plane with NNN kinetics.


\clearpage

\end{document}